\documentclass[useAMS,usenatbib,usegraphicx]{mn2e}
\usepackage{graphicx}
\usepackage[breaklinks,colorlinks,citecolor=black,linkcolor=black,urlcolor=black]{hyperref}

\newcommand*{\mysub}[2]{\ensuremath{#1_{\mathrm{#2}}}}
\newcommand*{\unit}[1]{\ensuremath{\mathrm{\, #1}}}
\newcommand*{\phmin}{\hspace{1.9ex}}

\newcommand*{\Zsun}{\ensuremath{\, Z_{\odot}}}

\newcommand*{\keV}{\unit{keV}}
\newcommand*{\erg}{\unit{erg}}
\newcommand*{\cm}{\unit{cm}}

\newcommand*{\second}{\unit{s}}

\newcommand*{\NH}{\mysub{N}{H}}
\newcommand*{\Mgas}{\mysub{M}{gas}}

\newcommand*{\fgas}{\mysub{f}{gas}}

\newcommand*{\rhocr}{\mysub{\rho}{cr}}

\newcommand*{\E}[1]{\ensuremath{\times 10^{#1}}}

\newcommand*{\ltsim}{\ {\raise-.75ex\hbox{$\buildrel<\over\sim$}}\ }
\newcommand*{\gtsim}{\ {\raise-.75ex\hbox{$\buildrel>\over\sim$}}\ }
\newcommand*{\proptosim}{\ {\raise-.75ex\hbox{$\buildrel\propto\over\sim$}}\ }

\newcommand*{\Chandra}{{\it Chandra}}
\newcommand*{\Suzaku}{{\it Suzaku}}
\newcommand*{\Planck}{{\it Planck}}

\newcommand*{\figscale}{0.9}

\begin{document}

\title[Evidence for Early Enrichment]{The Metallicity of the Intracluster Medium Over Cosmic Time: Further Evidence for Early Enrichment}

%\newauthor starts a new line
\author[A. B. Mantz et al.]{Adam B. Mantz,$^{1,2}$\thanks{E-mail: \href{mailto:amantz@slac.stanford.edu}{\tt amantz@slac.stanford.edu}} {}
  Steven W. Allen,$^{1,2,3}$
  R. Glenn Morris,$^{1,3}$
  Aurora Simionescu,$^4$\newauthor
  Ondrej Urban,$^{1,2}$
  Norbert Werner,$^{5,6,7}$
  Irina Zhuravleva$^{1,2}$
  \\$^1$Kavli Institute for Particle Astrophysics and Cosmology, Stanford University, 452 Lomita Mall, Stanford, CA 94305, USA
  \\$^2$Department of Physics, Stanford University, 382 Via Pueblo Mall, Stanford, CA 94305, USA
  \\$^3$SLAC National Accelerator Laboratory, 2575 Sand Hill Road, Menlo Park, CA  94025, USA
  \\$^4$Institute of Space and Astronautical Science (ISAS), JAXA, 3-1-1 Yoshinodai, Chuo-ku, Sagamihara, Kanagawa, 252-5210, Japan
  \\$^5$MTA-E\"otv\"os University Lend\"ulet Hot Universe Research Group, P\'azm\'any P\'eter s\'et\'any 1/A, Budapest, 1117, Hungary
  \\$^6$Department of Theoretical Physics and Astrophysics, Faculty of Science, Masaryk University, Kotl\'a\v{r}sk\'a 2, Brno, 611 37, Czech Republic
  \\$^7$School of Science, Hiroshima University, 1-3-1 Kagamiyama, Higashi-Hiroshima 739-8526, Japan
}
\date{Submitted 5 June 2017}

\pagerange{\pageref{firstpage}--\pageref{lastpage}} \pubyear{2017}
\maketitle
\label{firstpage}

\begin{abstract}
  We use \Chandra{} X-ray data to measure the metallicity of the intracluster medium (ICM) in 245 massive galaxy clusters selected from X-ray and Sunyaev-Zel'dovich (SZ) effect surveys, spanning redshifts $0<z<1.2$. Metallicities were measured in three different radial ranges, spanning cluster cores through their outskirts. We explore trends in these measurements as a function of cluster redshift, temperature, and surface brightness ``peakiness'' (a proxy for gas cooling efficiency in cluster centers). The data at large radii (0.5--1\,$r_{500}$) are consistent with a constant metallicity, while at intermediate radii (0.1--0.5\,$r_{500}$) we see a late-time increase in enrichment, consistent with the expected production and mixing of metals in cluster cores. In cluster centers, there are strong trends of metallicity with temperature and peakiness, reflecting enhanced metal production in the lowest-entropy gas. Within the cool-core/sharply peaked cluster population, there is a large intrinsic scatter in central metallicity and no overall evolution, indicating significant astrophysical variations in the efficiency of enrichment. The central metallicity in clusters with flat surface brightness profiles is lower, with a smaller intrinsic scatter, but increases towards lower redshifts. Our results are consistent with other recent measurements of ICM metallicity as a function of redshift. They reinforce the picture implied by observations of uniform metal distributions in the outskirts of nearby clusters, in which most of the enrichment of the ICM takes place before cluster formation, with significant later enrichment taking place only in cluster centers, as the stellar populations of the central galaxies evolve.
\end{abstract}

\begin{keywords}
  galaxies: clusters: intracluster medium -- X-rays: galaxies: clusters
\end{keywords}

\section{Introduction} \label{sec:intro}

The deep gravitational potential wells associated with clusters of galaxies retain nearly all of the baryonic matter involved in their formation or accreted later (e.g.\ \citealt*{Allen1103.4829}; \citealt{Borgani0906.4370}), including the metals produced in stars and ejected from galaxies into the ICM (e.g.\ \citealt{Boehringer2010A26ARv..18..127B}). The abundances and distribution of metals in the ICM thus encodes the history of star formation in cluster galaxies, as well as the processes that eject metals from the galaxies themselves and mix them with the surrounding gas.

It was shown early on that the ICM is enriched to about 0.3--0.5 of the Solar metallicity (\citealt{Mushotzky1978ApJ...225...21M}; we adopt the Solar relative abundance values of \citealt{Asplund0909.0948} throughout this work). Later observations with higher spatial resolution determined that the metallicity in nearby clusters is often centrally peaked, declining to $\sim0.3$ Solar outside of cluster centers \citep{Allen9802219, De-Grandi0310828, Leccardi0806.1445}. The most sensitive measurements of metallicity in the outskirts of galaxy clusters to date, extending to the virial radius, are from \Suzaku{} observations of nearby, X-ray bright clusters. \Suzaku{} Key Project observations of the outskirts of the Perseus cluster revealed a spatially homogenous metal distribution at $r>0.2\,r_{200}$, with an iron abundance with respect to Solar of $Z/\Zsun=0.314\pm0.012$ \citep{Werner1310.7948}. Consistent metallicity measurements have been obtained from \Suzaku{} observations of the outskirts of Coma \citep{Simionescu1302.4140}, Virgo \citep{Simionescu1506.06164}, and other nearby clusters \citep{Urban1706.01567} and groups \citep{Tholken1603.05255}. The uniformity and universality of the metallicity in cluster outskirts argues for early enrichment of the ICM, with most of the metals present being produced and mixed prior to the development of steep entropy gradients in clusters.

A natural consequence of this model is that, beyond their inner regions ($r\gtsim0.5\,r_{500}$), we should see no significant evolution in the metal content of clusters, at least out to redshifts $z\gtsim2$. Directly verifying this hypothesis is challenging, however, because the current X-ray observatories with sufficient spatial resolution to measure core-excluded metallicities in distant clusters suffer from higher backgrounds than \Suzaku{}, while using similar (relatively) low-energy-resolution CCD imaging spectrometers. Furthermore, cluster catalogs with available X-ray data that extend to high redshifts necessarily comprise the most massive clusters, with ICM temperatures $\gtsim3$\,keV; continuum emission is consequently significant, and only the Fe-K emission lines are typically detectable at CCD resolution. Consequently, the metallicity outside of the cores of cosmologically distant clusters is a low signal-to-noise observable, and will remain so pending the launch of new facilities with much larger collecting areas, good spatial resolution, and high energy resolution, such as {\it Athena} or {\it Lynx}. Previous studies of evolution in the metallicity of the ICM have yielded mixed results (e.g.\ \citealt{Balestra0609664, Maughan0703156, Anderson0904.1007}), although more recent results using XMM-{\it Newton} and \Chandra{} are consistent with little or no evolution \citep{Ettori1504.02107, McDonald1603.03035}.

Here we revisit this question using the largest cluster sample to date, 245 objects selected from X-ray and SZ surveys and observed by \Chandra{}, spanning redshifts  $0<z<1.2$. We perform this analysis for three radial ranges, with the smallest radii ($r<0.1\,r_{500}$) being dominated by cool cores when they exist, and the largest radii ($r>0.5\,r_{500}$) comfortably excluding the central gradients in metallicity previously observed. The latter measurement is challenging, but yields the most direct test of the predictions from \Suzaku{} regarding the metallicity in cluster outskirts. In Section~\ref{sec:data}, we introduce the \Chandra{} data set and the analysis methods used in this work. Section~\ref{sec:results} presents our results on the average cluster metallicity and its evolution as a function of cluster radius, temperature and morphology. We discuss the consequences of these results in Section~\ref{sec:discussion} and conclude in Section~\ref{sec:conclusion}. The appendices address the issue of fit statistics used for sparse X-ray spectra (the $C$ statistic versus $\chi^2$) and resulting biases (Appendix~\ref{sec:sims}), and the cross calibration of \Chandra{} and \Suzaku{} metallicity measurements (Appendix~\ref{sec:coma}).

\section{Data and Analysis} \label{sec:data}

\begin{figure}
  \centering
  \includegraphics[scale=\figscale]{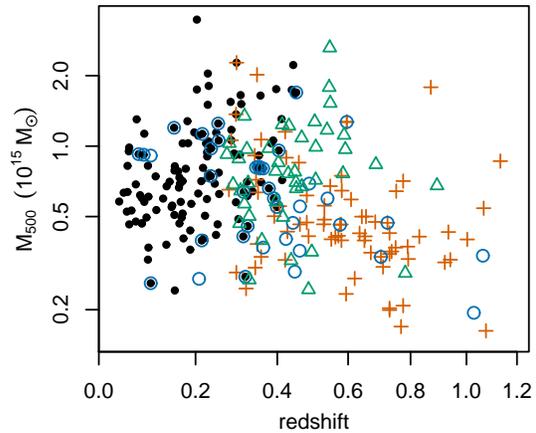}
  \caption{
    Mass--redshift distribution of our cluster data set, where we neglect error bars for clarity. Symbols are as follows: black, filled circles are selected from the X-ray flux limited BCS, REFLEX and bright MACS samples, with the further requirement that their 0.1--2.4\,keV RASS luminosities exceed $2.5\E{44}\erg\second^{-1}$ (see \citealt{Mantz1606.03407}); blue, empty circles correspond to massive, dynamically relaxed clusters identified from the Chandra archive; red crosses are clusters detected by SPT; and green triangles are X-ray selected clusters that do not fall into any of the other categories.
  }
  \label{fig:sample}
\end{figure}

The sample of clusters employed here consists of several differently selected subsamples, and  contains most of the known massive clusters that have been observed with \Chandra{}. Specifically, it contains clusters with \Chandra{} data that are either
\begin{enumerate}
\item used in studies of cluster cosmology and scaling relations by \citet{Mantz1407.4516, Mantz1606.03407}. This subsample is formed from the ROSAT All-Sky Survey (RASS) based BCS \citep{Ebeling1998MNRAS.301..881E}, REFLEX \citep{Bohringer0405546}, and Bright MACS \citep{Ebeling1004.4683} catalogs, with a minimum 0.1--2.4\,keV luminosity (as measured from RASS) of $2.5\E{44}\erg\second^{-1}$. 
\item identified as massive and dynamically relaxed based on a morphological search of the \Chandra{} archive (see \citealt{Mantz1402.6212, Mantz1502.06020}).
\item detected through their SZ effect by the South Pole Telescope (SPT; \citealt{Bleem1409.0850}). Note that \Chandra{} follow-up preferentially targeted the most massive SPT clusters.
\item X-ray selected from the Faint MACS catalog \citep{Ebeling0703394, Mann1111.2396}, the BCS (extending to a lower flux limit than was used by \citealt{Mantz1407.4516}), and from earlier samples of relaxed clusters (in particular, that of \citealt{Allen0706.0033}).
\end{enumerate}
In addition, we require the temperature measured from our analysis, below, at intermediate radii (0.1--0.5\,$r_{500}$) to be $\geq5$\,keV, in order to ensure that the sample consists of genuinely massive clusters (this cut only eliminates 14 clusters from the sample, given the initial selection). The \Chandra{} data for all clusters in the sample have been analyzed uniformly, as described below. Figure~\ref{fig:sample} shows the distribution in redshift and mass of the different subsamples.

Our reduction of the \Chandra{} data follows the procedure detailed by \citet{Mantz1502.06020}, with the exception that this work employs a more recent version of the \Chandra{} calibration files (specifically, {\sc caldb}\footnote{\url{http://cxc.harvard.edu/caldb/}} version 4.6.7). In brief, the raw data were reprocessed to produce level 2 event files, and were filtered to eliminate periods of high background. For each observation, a corresponding quiescent background data set was produced using the \Chandra{} blank-sky data,\footnote{\url{http://cxc.cfa.harvard.edu/ciao/threads/acisbackground/}} rescaled according to the measured count rate in the 9.5--12\,keV band. Each cluster field was then tested for the presence of a soft Galactic foreground, which is absent from the blank-sky data sets, as described in \citet{Mantz1402.6212}. When present, this foreground component was constrained simultaneously with the cluster model in our subsequent spectral analysis.

A cluster's mass and its associated characteristic radius are jointly defined in terms of the critical density at its redshift, $M_\Delta = (4/3) \pi \Delta \rhocr(z) r_\Delta^3$, for the conventional ``overdensity'' of $\Delta=500$. In this work, we are only interested in obtaining estimates of $r_{500}$ for each cluster so that constraints on the metallicity in comparable regions can be extracted. We obtained these $r_{500}$ estimates following the procedure described by \citet{Mantz1606.03407}, which is summarized below. After defining the center of the X-ray emission, spectra were extracted in concentric annuli, binned to have at least 1 count per channel, and were used to fit in {\sc xspec}\footnote{\url{http://heasarc.gsfc.nasa.gov/docs/xanadu/xspec/}} a non-parametric, spherically symmetric model for the three-dimensional gas density in a given cluster. Cluster emission was modeled using the {\sc apec} plasma model ({\sc atomdb} version 2.0.2). Relative metal abundances were fixed to the solar ratios of \citet{Asplund0909.0948}, with the overall metallicity allowed to vary. Photoelectric absorption by Galactic gas was accounted for using the {\sc phabs} model, employing the cross sections of \citet{Balucinska1992ApJ...400..699B}. For each cluster field, the equivalent absorbing hydrogen column densities, \NH{}, were fixed to the values from the H{\sc i} survey of \citet{Kalberla0504140}, provided that the published values were $<10^{21}\cm^{-2}$. When the published column densities were $\geq10^{21}\cm^{-2}$, we included \NH{} as a free parameter fitted to the X-ray data. The likelihood of spectral models was evaluated using the \citet{Cash1979ApJ...228..939} statistic, as modified by Arnaud et~al.\footnote{XSPEC Users' Guide, Appendix B: \url{https://heasarc.gsfc.nasa.gov/docs/xanadu/xspec/manual/manual.html}} (the $C$-statistic) to account for the use of an empirical background model (i.e.\ the blank-sky data). Confidence regions were determined by Markov Chain Monte Carlo (MCMC) explorations of the parameter space using the {\sc lmc} code.\footnote{\url{https://github.com/abmantz/lmc}} From the resulting constraints on the enclosed gas mass as a function of radius, we can obtain an estimate of $r_{500}$, based on the integrated gas mass fraction of $\fgas(r_{500})=0.125$ measured by \citet{Mantz1606.03407}, by finding the radius that satisfies
\begin{equation} \label{eq:rDelta}
  \Mgas(r) = \frac{4\pi}{3} \, 500\rhocr(z) \fgas(r_{500}) r^3.
\end{equation}

Given an estimate of $r_{500}$ for each cluster, we extract new spectra in each of three annular apertures, spanning radii of 0.0--0.1\,$r_{500}$, 0.1--0.5\,$r_{500}$ and 0.5--1.0\,$r_{500}$. Each spectrum is fitted independently using the spectral analysis described above. In this case, however, the cluster emission in each region is modeled by a single {\sc apec} emission component with free temperature, metallicity and normalization parameters; Galactic absorption and foregrounds (when present) are handled as before. We allow the temperature, $kT$, and metallicity, $Z$, of the cluster model to vary over the ranges 0.1--64\,keV and 0.0--5.0$\Zsun$, respectively. Because the apertures used in these fits are defined to cover comparable regions of each cluster rather than being based on signal-to-noise considerations, some cluster spectra are inadequate to constrain the model. Such unconstrained fits could potentially bias our results, given the uniform and asymmetric nature of the priors on $kT$ and $Z$ relative to typical values of $\sim6\keV$ and $\sim0.3\Zsun$, even though clusters with large uncertainties would naturally be down-weighted in our subsequent analysis. In practice, we find that unconstrained spectral fits can be automatically identified by applying generous cuts on the recovered parameters, specifically by requiring values of $kT>30\keV$ and $Z>2\Zsun$ to be ruled out at the 95 per cent confidence level.\footnote{The only exception to this rule is CL\,J1415+3612, which in the 0.0--0.1\,$r_{500}$ region has $kT=(6.1\pm0.4)\keV$ and $Z=(1.9\pm0.4)\Zsun$ (see also \citealt{Santos1111.3642}). Even given the large cluster-to-cluster scatter in this central region (Section~\ref{sec:small}), this value is a sufficiently extreme outlier that we remove it from subsequent analysis.} Reasonable adjustments of these cuts have no impact on our results, due to the fact that the few measurements greater or fewer that are excluded have little statistical leverage. At the lowest redshifts, the limited \Chandra{} field of view also sometimes prevents the larger apertures from being included in our analysis at all.

\begin{figure*}
  \centering
  \includegraphics[scale=\figscale]{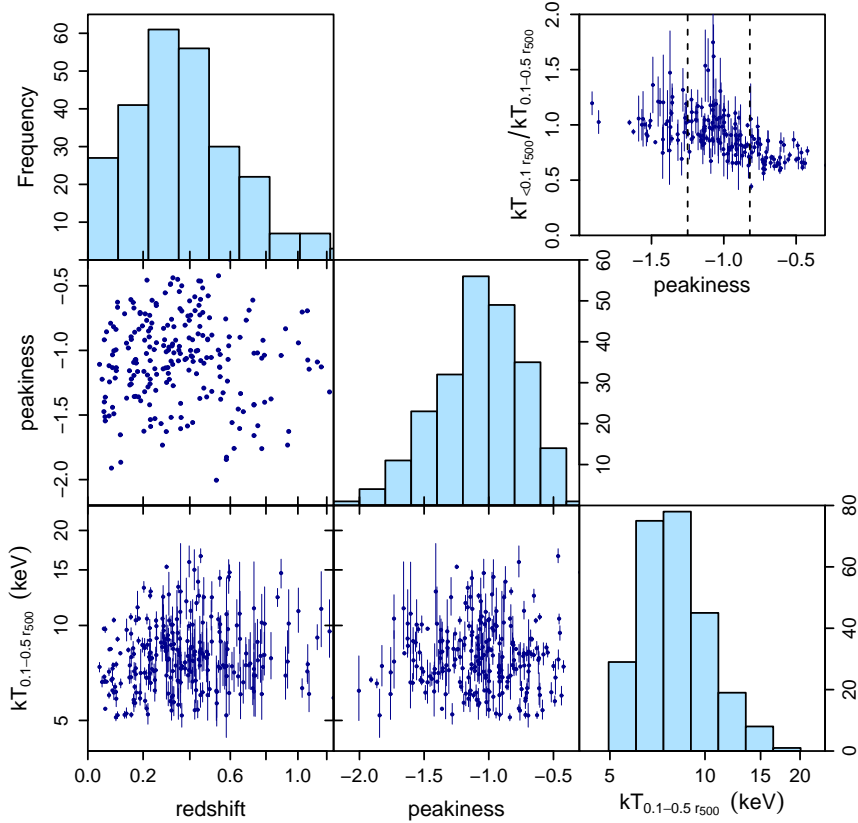}
  \caption{
    Properties of our cluster sample. The lower-left triangle shows how the clusters are distributed in terms of redshift, temperature at intermediate radii, and surface brightness peakiness, including histograms of these parameters. The upper-right panel shows peakiness against the temperature in cluster centers, displaying the expected trend wherein cool-core clusters tend to have sharper surface brightness peaks. The vertical, dashed lines divide the sample into 3 peakiness ranges that we refer to in this analysis.
  }
  \label{fig:tri}
\end{figure*}

In addition to redshift and X-ray spectral measurements, we characterize the clusters in our sample morphologically, using measurements of the surface brightness `peakiness' from \citet{Mantz1502.06020}. This metric is a proxy for the `cool cores' in clusters, i.e.\ the presence of significant density enhancements and temperature decrements that appear in the centers of some clusters. The upper-right panel of Figure~\ref{fig:tri} explicitly verifies this relationship, showing peakiness against the ratio of the temperatures measured in the 0.0--0.1\,$r_{500}$ and 0.1--0.5\,$r_{500}$ apertures. The lower-left triangle portion of the figure shows the distribution of our cluster sample in redshift, peakiness and the temperature at intermediate radii (0.1--0.5\,$r_{500}$), along with histograms of these parameters.

 In Section~\ref{sec:results}, we report the results of fitting a model for the mean cluster metallicity as a function of redshift and temperature. Simple curve fitting techniques such as least squares are inadequate for this task, due to the non-zero measurement covariance between $kT$ and $Z$ measured from the same spectrum, and the often non-Gaussian shape of the posterior distribution of $Z$ (resulting from the prior cut at $Z=0$). We therefore make use of the full set of MCMC samples for each cluster when fitting these higher-level (hierarchical) models. In detail, this means that the samples for each cluster are importance weighted according to the hierarchical model (e.g.\ a mean $Z$ and log-normal intrinsic scatter) and then numerically integrated to provide a final posterior probability. The sampling itself was performed using {\sc rgw},\footnote{\url{https://github.com/abmantz/rgw/}} which implements the algorithm of \citet{Goodman10.2140-camcos.2010.5.65} in the {\sc r} environment.\footnote{\url{https://www.r-project.org}}

\section{Results} \label{sec:results}

\begin{figure*}
  \centering
  \includegraphics[scale=1]{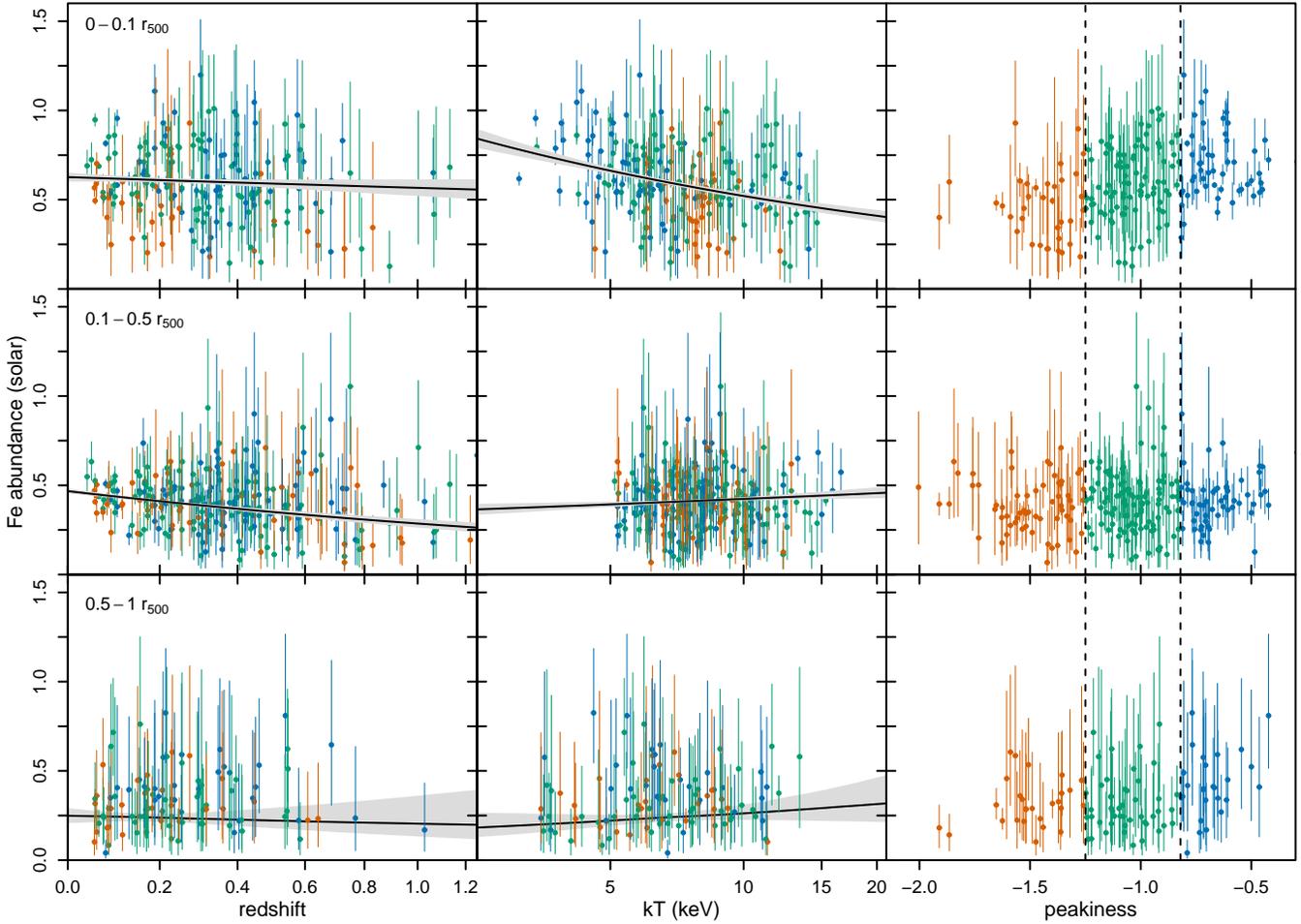}
  \caption{
    Measured metallicities in three apertures (rows) are plotted as a function of redshift, temperature, and peakiness (columns). Points are color coded according to peakiness (right column) and correspond to the posterior median for each cluster; error bars span the 15.85th--84.15th percentiles (68.3 per cent confidence). In the left and center columns,  black lines and gray shading similarly show the median and 68.3 per cent confidence region for the mean metallicity power-laws with $1+z$ or $kT$ (Equation~\ref{eq:Zmodel}), evaluated at the pivot temperature or redshift. In the right column, vertical, dashed lines show the thresholds used to define subsamples of clusters based on peakiness.
  }
  \label{fig:metalz}
  \label{fig:metalT}
  \label{fig:peakiness}
\end{figure*}

Our measurements of metallicity in each aperture are plotted against redshift, temperature in the corresponding region, and peakiness in Figure~\ref{fig:metalz}. By eye, the clearest trends are between the central (0.0--0.1\,$r_{500}$) metallicity, temperature and peakiness. However, these trends are not independent, due to the relationship between peakiness and the central temperature (top-right panel of Figure~\ref{fig:tri}, top-center panel of Figure~\ref{fig:metalT}). Consequently, we consider models where metallicity varies with redshift and temperature, but do not explicitly include peakiness as an additional explanatory variable. However, in subsequent discussion, we will contrast subsamples of clusters in 3 categories: those with peakiness values of $p\geq-0.82$, $-1.25\leq p<-0.82$, and $p<-1.25$. Henceforth, we refer to these groups as high (H), medium (M), and low-peakiness (L) clusters.

To be precise, we fit the model
\begin{equation} \label{eq:Zmodel}
  Z = Z_0 \left(\frac{1+z}{1+z_\mathrm{piv}}\right)^{\beta_{1+z}} \left(\frac{kT}{kT_\mathrm{piv}}\right)^{\beta_{kT}},
\end{equation}
including a log-normal intrinsic scatter, $\sigma_{\ln Z}$, where $Z$ and $kT$ apply to a particular aperture. The pivot redshift and temperature, $z_\mathrm{piv}$ and $kT_\mathrm{piv}$, can be chosen in each case to minimize the posterior correlation of $Z_0$ with $\beta_{1+z}$ and $\beta_{kT}$. Table~\ref{tab:fit} shows the parameter constraints obtained by fitting the \Chandra{} measurements corresponding to each aperture (without selecting based on peakiness); 68.3 per cent confidence  bands for the mean metallicity trends are shown in Figure~\ref{fig:metalz}. Results from fitting the 3 peakiness-based subsets of clusters in each aperture appear in Table~\ref{tab:fitpeak}. In the following subsections, we discuss in more detail the results for each aperture.

\begin{table*}
  \begin{center}
    \caption{
      Constraints on the model of Equation~\ref{eq:Zmodel} in three apertures from our \Chandra{} data. [1] Radial range in units of $r_{500}$; [2--3] pivot redshift and temperature (in keV) corresponding to the normalization, $Z_0$; [4] number of clusters contributing to the fit for a given aperture; [5--8] best-fitting values (posterior modes) and 68.3 per cent maximum-probability confidence intervals for the model parameters. 
    }
    \label{tab:fit}
    \vspace{-1ex}
    \begin{tabular}{cccccccc}
      \hline
      Aperture & $z_\mathrm{piv}$ & $kT_\mathrm{piv}$ & $N_\mathrm{cl}$ & $Z_0/\Zsun$ & \phmin$\beta_{1+z}$ & \phmin$\beta_{kT}$ & $\sigma_{\ln Z}$\\
      \hline
      0.0--0.1 & 0.23 & 6.4 & 186 & $0.607\pm0.012$ & $-0.14\pm0.17$ & $-0.35\pm0.06$ & $0.18\pm0.02$\\
      0.1--0.5 & 0.19 & 8.0 & 245 & $0.413\pm0.007$ & $-0.71\pm0.15$ & \phmin$0.10\pm0.07$ & $0.08\pm0.02$\\
      0.5--1.0 & 0.17 & 6.7 & 86 & $0.240\pm0.022$ & $-0.30\pm0.91$ & \phmin$0.22\pm0.34$ & $0.00^{+0.17}_{-0.00}$\\
      \hline
   \end{tabular}
  \end{center}
\end{table*} 

\begin{table*}
  \begin{center}
    \caption{
      Constraints on the model of Equation~\ref{eq:Zmodel} in three apertures and for peakiness-based subsets of the cluster sample from our \Chandra{} data. Columns are as in Table~\ref{tab:fit}, except that the first column here indicates whether the constraints apply to high (H: $p\geq-0.82$), medium (M: $-1.25\leq p<-0.82$), or low (L: $p<-1.25$) peakiness clusters.
    }
    \label{tab:fitpeak}
    \vspace{-1ex}
    \begin{tabular}{ccccccccc}
      \hline
      Peakiness & Aperture & $z_\mathrm{piv}$ & $kT_\mathrm{piv}$ & $N_\mathrm{cl}$ & $Z_0/\Zsun$ & \phmin$\beta_{1+z}$ & \phmin$\beta_{kT}$ & $\sigma_{\ln Z}$\\
      \hline
      H & 0.0--0.1 & 0.31 & 5.5 & 53 & $0.652\pm0.022$ & $-0.08\pm0.28$ & $-0.24\pm0.13$ & $0.19\pm0.03$\\
      H & 0.1--0.5 & 0.24 & 8.2 & 56 & $0.421\pm0.013$ & $-0.44\pm0.26$ & \phmin$0.40\pm0.14$ & $0.00^{+0.10}_{-0.00}$\\
      H & 0.5--1.0 & 0.21 & 5.9 & 23 & $0.370\pm0.080$ & $-0.38^{+1.31}_{-2.05}$ & $-0.81^{+0.56}_{-1.07}$ & $0.00^{+0.19}_{-0.00}$\medskip\\
      M & 0.0--0.1 & 0.19 & 7.1 & 90 & $0.614\pm0.018$ & $-0.45\pm0.30$ & $-0.30\pm0.09$ & $0.15\pm0.03$\\
      M & 0.1--0.5 & 0.18 & 8.1 & 107 & $0.426\pm0.009$ & $-0.89^{+0.15}_{-0.47}$ & \phmin$0.08\pm0.10$ & $0.07\pm0.03$\\
      M & 0.5--1.0 & 0.16 & 7.2 & 35 & $0.199\pm0.037$ & \phmin$0.66\pm2.19$ & \phmin$0.73\pm0.67$ & $0.00^{+0.21}_{-0.00}$\medskip\\
      L & 0.0--0.1 & 0.11 & 7.6 & 33 & $0.489\pm0.024$ & $-2.07\pm0.79$ & $-0.41\pm0.34$ & $0.09\pm0.05$\\
      L & 0.1--0.5 & 0.14 & 7.6 & 58 & $0.388\pm0.012$ & $-0.97\pm0.33$ & $-0.25\pm0.19$ & $0.00^{+0.08}_{-0.00}$\\
      L & 0.5--1.0 & 0.13 & 6.8 & 25 & $0.245\pm0.037$ & $-1.58\pm1.64$ & \phmin$0.67\pm0.68$ & $0.00^{+0.16}_{-0.00}$\\
      \hline
   \end{tabular}
  \end{center}
\end{table*}

\subsection{Small Radii (0.0--0.1\,$r_{500}$)} \label{sec:small}

In cluster centers, a trend towards higher metallicity in cool/peaky cores is clear ($\beta_{kT}$ is almost $6\sigma$ different from zero in Table~\ref{tab:fit}). Controlling for this trend, our measurements are consistent with a constant value with redshift, although with a significant intrinsic scatter ($0.18\pm0.02$ in $\ln Z$). Breaking these features down in terms of peakiness, we see that the intrinsic scatter and temperature-dependence are significantly non-zero only for the H and M subsamples (Table~\ref{tab:fitpeak}). For the L subsample, the scatter and temperature dependence are both consistent with zero. As a function of redshift, the H sample is consistent with a constant, while the L sample evolves strongly, with the M sample presenting an intermediate case (a $\sim1.5\sigma$ departure from zero).

The high metallicity and lack of evolution in the centers of cool-core clusters (H), and the correlation with peakiness, suggest substantial early enrichment of the coolest, lowest-entropy gas through vigorous star formation, with subsequent enrichment by aging stellar populations playing a relatively minor role. The large scatter within this population implies that properties of the clusters beyond their redshifts and temperatures, perhaps related to accretion rates and the strength of AGN feedback, can also have a significant influence on central metallicities. In clusters without cool cores (L), the scatter in central metallicity and its dependence on central temperature are minimal. At the same time, this cluster population steadily accumulates metals in cluster centers, presumably due to the continuing evolution and mass loss of stellar populations in the central galaxies. Clusters with medium-strength cores (M) lie between these two extremes.

\begin{table*}
  \begin{center}
    \caption{
      Constraints on the model of Equation~\ref{eq:Zmodel} in the 0.1--0.5\,$r_{500}$ aperture from our \Chandra{} data, for different cluster selections. The first column indicates the selection, which is either a redshift range or `PSZ' (see text). Otherwise, columns are as in Table~\ref{tab:fit}.
    }
    \label{tab:fit5}
    \vspace{-1ex}
    \begin{tabular}{cccccccc}
      \hline
      Selection & $z_\mathrm{piv}$ & $kT_\mathrm{piv}$ & $N_\mathrm{cl}$ & $Z_0/\Zsun$ & \phmin$\beta_{1+z}$ & \phmin$\beta_{kT}$ & $\sigma_{\ln Z}$\\
      \hline
      $z<0.4$ & 0.16 & 7.9 & 149 & $0.423\pm0.007$ & $-0.53\pm0.23$ & $0.08\pm0.08$ & $0.08\pm0.02$\\
      $z\ge0.4$ & 0.55 & 8.7 & 99 & $0.327\pm0.017$ & $-0.23\pm0.56$ & $0.20\pm0.21$ & $0.00^{+0.14}_{-0.00}$\\
      PSZ & 0.13 & 8.3 & 79 & $0.428\pm0.008$ & $-0.70\pm0.39$ & $0.10\pm0.11$ & $0.07\pm0.02$\\
      \hline
   \end{tabular}
  \end{center}
\end{table*}

\subsection{Intermediate Radii (0.1--0.5\,$r_{500}$)} \label{sec:intermediate}

Metallicities measured at these radii lack the clear dependence on the local temperature that is present in cluster centers. They do, however, strongly prefer evolving models, irrespective of peakiness ($\beta_{1+z}=-0.71\pm0.15$ overall). Similarly, the intrinsic scatter is small, $\ltsim10$ per cent, for all three peakiness-based subsamples.

\begin{figure}
  \centering
  \includegraphics[scale=\figscale]{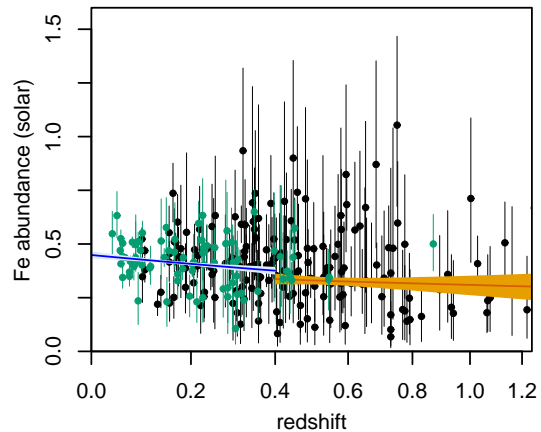}
  \caption{
    Measured metallicities in the 0.1--0.5\,$r_{500}$ aperture, with green points indicating those detected by \Planck{} with signal-to-noise $>10$. Blue and orange shading show the 68.3 per cent confidence regions for power-law evolution models fit to the data (irrespective of \Planck{} selection) at $z<0.4$ and $z\ge0.4$, respectively.
  }
  \label{fig:metal5}
\end{figure}

However, the evidence for evolution does not come uniformly from the data at all redshifts. Splitting the sample at $z=0.4$ (approximately the mean), we find that the $z<0.4$ sample prefers evolution at $\sim2\sigma$ significance, while the $z\ge0.4$ sample is consistent with a constant (Table~\ref{tab:fit5}). The latter, however, has larger statistical uncertainties, and does not rule out the evolution implied by the lower-redshift sample. Figure~\ref{fig:metal5} visualizes these fit results.

Referring to Figure~\ref{fig:sample}, we see that clusters in our sample at $z<0.4$ are predominantly X-ray selected (from RASS), while the majority at $z>0.4$ are SZ selected (from SPT). The possibility therefore exists that the different evolutionary behavior seen at $z<0.4$ and $z>0.4$ is due to selection effects. To test this, we take advantage of the fact that the \Planck{} SZ-selected cluster catalog overlaps substantially with both the RASS and SPT catalogs \citep{Planck1502.01598}. Table~\ref{tab:fit5} shows results from fitting a subset of our data set from the \Planck{}-SZ catalog (PSZ), with SZ signal-to-noise $>10$ (highlighted in Figure~\ref{fig:metal5}). This signal-to-noise threshold was chosen to provide a sample of comparable size to the others considered here, but unfortunately comes at the expense of relatively low completeness (65 per cent). The fit to this PSZ subsample, which spans $0<z<0.87$, is consistent with the full data set, with $\beta_{1+z}=-0.70\pm0.39$.

Thus, while the data are not entirely conclusive, the indications point towards a constant metallicity at radii of 0.1--0.5\,$r_{500}$ at high redshift, with metallicities beginning to rise at late times, independent of selection effects. Interestingly, the metallicity evolution at these radii as a function of peakiness is qualitatively similar to that seen in cluster centers (i.e.\ stronger evidence of evolution in less peaky clusters). In situ enrichment from stellar evolution in cluster galaxies provides a plausible mechanism in both cases, although the gas at radii of 0.1--0.5\,$r_{500}$ may also be enriched through mixing with gas from cluster centers, driven by mergers or AGN outbursts.

\subsection{Large Radii (0.5--1.0\,$r_{500}$)} \label{sec:outskirts}

Only about 35 per cent of clusters in our data set provide reliable metallicity constraints at these radii, and the statistical uncertainties are relatively large; consequently, the model parameters are much less well constrained here than in the other apertures. The \Chandra{} data are consistent with a constant metallicity as a function of redshift and temperature ($\beta_{1+z}=-0.30\pm0.91$ and $\beta_{kT}=0.22\pm0.34$), with no intrinsic scatter ($\sigma_{\ln Z}=0.00^{+0.17}_{-0.00}$). Note that the a posteriori correlation of the scatter with the slope parameters is small, such that the upper limit on the scatter does not change if the metallicity is assumed to be constant with redshift and temperature.

Our results for this aperture can be compared with \Suzaku{} measurements of metallicity at similar radii in nearby clusters. The low particle background of \Suzaku{} makes it ideal for probing the faint ICM in cluster outskirts, while its large point spread function and concerns about scattered light effectively restrict its usage to radii $\gtsim0.5\,r_{500}$ and cluster redshifts $\ltsim0.2$. Here we make use of measurements made from very deep observations of the Coma and Perseus clusters (\citealt{Simionescu1302.4140, Werner1310.7948}; redshifts of 0.023 and 0.018, respectively), as well as metallicity measurements for 5 clusters at redshifts $0.063<z<0.183$ by \citet{Urban1706.01567}, all of which satisfy the same temperature threshold as our \Chandra{} data set ($kT>5$\,keV). Table~\ref{tab:suzaku} lists these \Suzaku{} metallicity measurements.

\begin{table}
  \begin{center}
    \caption{
      Redshifts and metallicities measured from \Suzaku{} data by \citet{Werner1310.7948}, \citet{Simionescu1302.4140} and \citet{Urban1706.01567} for hot, low-redshift clusters ($kT>5$\,keV, $z<0.2$). The values listed here are relative to the \citet{Asplund0909.0948} Solar abundances, and represent the average of the available measurements at cluster radii $\gtsim0.5\,r_{500}$. The \Suzaku{} metallicity profiles typically extend to $r>r_{500}$, and are consistent with a constant value over the range averaged.
    }
    \label{tab:suzaku}
    \vspace{-1ex}
    \begin{tabular}{lcr@{ $\pm$ }l}
      \hline 
      Cluster & $z$ & \multicolumn{2}{c}{$Z^\mathrm{Suz}/\Zsun$} \\
      \hline 
      Perseus & 0.018  &  0.314  &  0.012  \\
      Coma & 0.023  &  0.291  &  0.037  \\
      Abell 1795 & 0.063  &  0.307  &  0.025  \\
      Abell 2029 & 0.078  &  0.326  &  0.065  \\
      Abell 2142 & 0.089  &  0.357  &  0.035  \\
      Abell 2204 & 0.152  &  0.396  &  0.069  \\
      Abell 1689 & 0.183  &  0.353  &  0.131  \\
      \hline
   \end{tabular}
  \end{center}
\end{table}

\begin{figure}
  \centering
  \includegraphics[scale=\figscale]{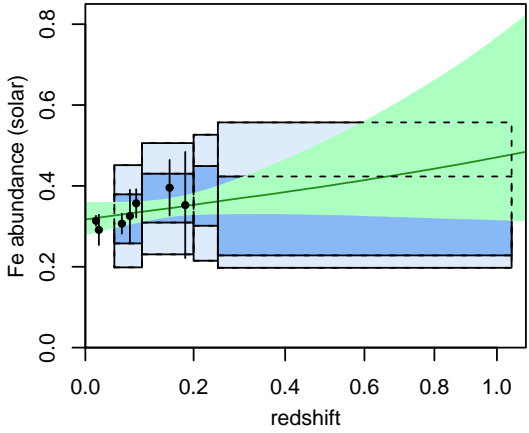}
  \caption{
    Measured metallicities at $r>0.5\,r_{500}$ from \Suzaku{} (black points). For visualization purposes, \Chandra{} measurements in the 0.5--1.0\,$r_{500}$ aperture are represented by blue-shaded boxes; these show the 68.3 and 95.4 per cent confidence intervals for the $Z_0$ parameter when fitting clusters in each redshift bin (fixing $\beta_{1+z}=0$), and are corrected by the best-fitting \Suzaku{}/\Chandra{} cross calibration factor (37.5 per cent). Green shading shows the 68.3 per cent confidence regions for power-law evolution models fit to the combined data set.
  }
  \label{fig:metal4}
\end{figure}

While incorporating the \Suzaku{} measurements into our evolution analysis can in principle yield tighter constraints, an overall offset cross-calibrating Fe abundance measurements with the two telescopes must be accounted for. The need for such an offset is clear from a comparison of the average metallicity measured at large radii from the two data sets: $0.240\pm0.022$ from \Chandra{} and $0.314\pm0.012$ (Perseus; \citealt{Werner1310.7948}) or $0.316\pm0.012$ (other clusters; \citealt{Urban1706.01567}) from \Suzaku{}. Note that the pivot redshift for our \Chandra{} data in this aperture is relatively low (0.17), making it unlikely that this disagreement reflects true evolution of the type we are aiming to measure (see also Figure~\ref{fig:metal4}). A comparable offset is found by directly comparing \Chandra{} and \Suzaku{} measurements at large radii in the Coma cluster (see Section~\ref{sec:systematics} and Appendix~\ref{sec:coma}). We include an overall cross-calibration factor as a free parameter of the model, adopting a Gaussian prior of $\ln(Z^\mathrm{Suz}/Z^\mathrm{Cha})=0.41\pm0.14$ based on the Coma data.

Given the consistency of the \Chandra{} data with a temperature-independent metallicity at these radii, we simplify the joint \Chandra+\Suzaku{} analysis by assuming $\beta_{kT}=0$. Where measurements are available from both \Chandra{} and \Suzaku{}, we use the \Suzaku{} data preferentially (Coma is not explicitly included in the fit, as it provides the basis of the cross-calibration prior above). The constraints obtained from this combined data set are summarized in Table~\ref{tab:fit4}, which lists $Z_0$ with and without the cross-calibration applied. The constraints on evolution, $\beta_{1+z}=0.35^{+1.18}_{-0.36}$, are slightly tighter than those from \Chandra{} alone, and are consistent with a constant value; negative values of $\beta_{1+z}$ in particular are less favored. Despite the high precision of the \Suzaku{} measurements, the intrinsic scatter remains consistent with zero. Our a posteriori constraint on the cross-calibration parameter is $\ln(Z^\mathrm{Suz}/Z^\mathrm{Cha})=0.375\pm0.096$, somewhat tighter than, and consistent with, the prior from Coma.

\begin{table}
  \begin{center}
    \caption{
      Constraints on the model of Equation~\ref{eq:Zmodel} (with $\beta_{kT}$ fixed to zero) in the 0.5--1.0\,$r_{500}$ aperture from the combination of our \Chandra{} and \Suzaku{} data. The $Z_0$ parameter is given relative to both the \Chandra{} and \Suzaku{} references.
    }
    \label{tab:fit4}
    \vspace{-1ex}
    \begin{tabular}{lc}
      \hline
      $z_\mathrm{piv}$ & 0.15 \\
      $N_\mathrm{cl}$ & 89 \\
      $Z_0^\mathrm{Cha}/\Zsun$ & $0.234\pm0.016$ \\
      $Z_0^\mathrm{Suz}/\Zsun$ & $0.342\pm0.025$ \\
      $\beta_{1+z}$ & $0.35^{+1.18}_{-0.36}$ \\
      $\sigma_{\ln Z}$ & $0.00^{+0.06}_{-0.00}$ \\
      $\ln(Z^\mathrm{Suz}/Z^\mathrm{Cha})$ & $0.375\pm0.096$ \\
      \hline
   \end{tabular}
  \end{center}
\end{table}

Figure~\ref{fig:metal4} shows the \Suzaku{} data set, along with the 68.3 per cent confidence limits for the model. For visualization, the \Chandra{} data are represented by their 68.3 and 95.4 per cent confidence intervals for the $Z_0$ parameter (corrected to the \Suzaku{} reference) when the model in Equation~\ref{eq:Zmodel} is fit to clusters in the redshift bins shown (bounded by $z=0.05$, 0.1, 0.2, 0.25 and 1.05) independently (fixing $\beta_{1+z}=0$). 

\subsection{Synthesis}

\begin{figure}
  \centering
  \includegraphics[scale=1]{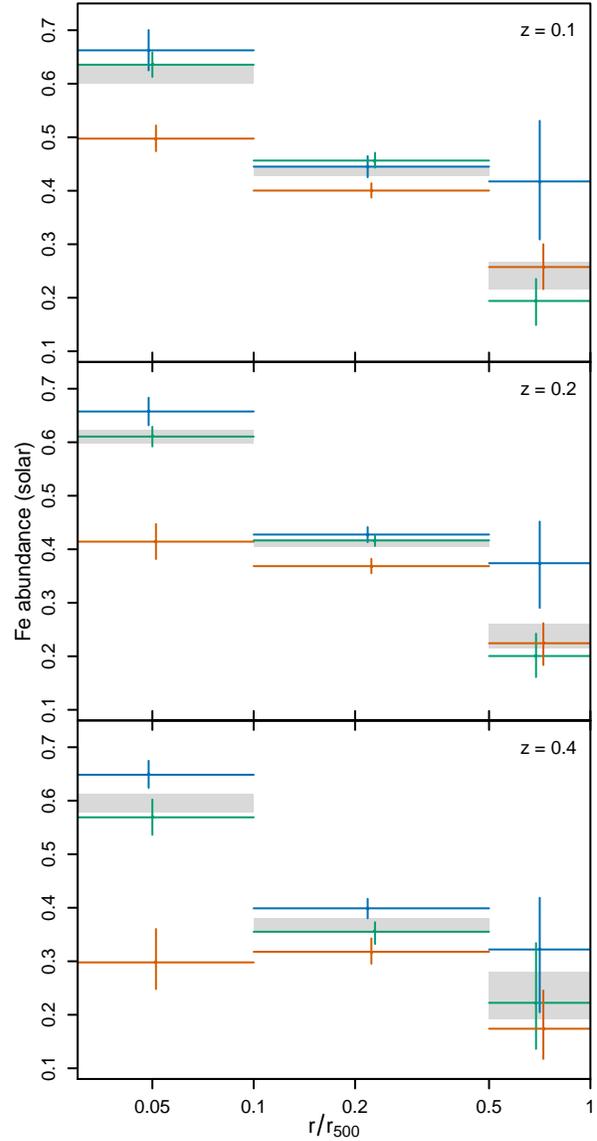}
  \caption{
    Metallicity profiles our constraints on the model in Equation~\ref{eq:Zmodel} at redshifts 0.1, 0.2 and 0.4. Gray shading shows the 68.3 per cent confidence region at each redshift and radial aperture from fitting the full cluster sample, and blue, green and red crosses respectively show the same for the H, M and L subsamples. For each cluster sample and aperture, the pivot temperature (i.e.\ a typical temperature for the clusters in the fit) is adopted when evaluating the model.
  }
  \label{fig:profile}
\end{figure}

The results in Table~\ref{tab:fitpeak} are summarized visually in Figure~\ref{fig:profile}, which shows metallicity constraints in each of the apertures considered here, for H, M and L peakiness clusters, at three different redshifts spanning the range that is well sampled by our data set ($z=0.1$, 0.2 and 0.4). Here we can see the broad trends with redshift and morphology identified in the previous sections. The profiles for H and M clusters are relatively constant with redshift, excepting the intermediate-radius aperture of M clusters, where the metallicity does increase with time. In general, the M clusters agree well with the full-sample average profile, with H clusters being marginally higher in metallicity. In contrast, at radii $<0.5\,r_{500}$, the L clusters have significantly lower metallicities than the average, particularly in cluster centers ($r<0.1\,r_{500}$). However, the $r<0.5\,r_{500}$ metallicity in L clusters evolves relatively strongly as well, approaching the profiles of the other subsamples at low redshifts. At large radii, $r>0.5\,r_{500}$, the three subsamples of clusters are consistent with one another, and with a constant value (full-sample intrinsic scatter of $\sigma_{\ln Z}=0.00^{+0.17}_{-0.00}$).

\subsection{Systematic Checks} \label{sec:systematics}

In this section, we consider the impact of a number of potential systematic issues on our measurements. To begin with, we tested on simulated data the ability of our spectral fitting procedure to recover unbiased posterior distributions for temperature and metallicity as a function of signal-to-noise. Specifically, these tests apply to the use of the modified $C$ statistic when spectra are grouped to $\geq1$ count per channel, using a realistic \Chandra{} background and for redshifts and temperatures typical of our data set. Details are presented in Appendix~\ref{sec:sims}. To summarize, we find that our approach produces accurate estimates and confidence intervals. This is in contrast to the still widely used practice of binning spectra more heavily and using the $\chi^2$ fit statistic, which generally produces biased temperature and metallicity estimates over a range of typical exposure times (see also \citealt{Leccardi0705.4199, Humphrey0811.2796}). Within the \Chandra{} data themselves, there is no overall trend of measured metallicity with signal-to-noise (measured in an energy band that excludes the Fe-K line).

Several more tests yielded null results. These were: shifting the normalizations of the blank-sky backgrounds by $\pm10$ per cent; fitting two-temperature models, with common metallicity and the second temperature fixed to 0.6\,keV; and using the {\sc mekal} thermal emission model instead of {\sc apec}. Each of these adjustments resulted in shifts in the metallicity measurements for individual clusters, but the shifts were generally within the $1\sigma$ statistical uncertainties, and none of these tests impact our conclusions as a function of redshift. Our \Chandra{} results thus appear to be internally robust against reasonable uncertainties in the X-ray background, the temperature structure of the ICM, and modelling uncertainties.

The remaining, and significant, question is whether there may be an overall bias in metallicities measured with \Chandra{}, in particular with respect to the \Suzaku{} measurements that we compare to in Section~\ref{sec:outskirts}. Comparison of Tables~\ref{tab:fit} and \ref{tab:suzaku} suggests that the average metallicity at $r>0.5\,r_{500}$ measured from the two telescopes for clusters at the same redshifts differs substantially (i.e.\ a substantial correction is needed to produce the agreement shown in Figure~\ref{fig:metal4}). Ideally, any cross-calibration factor could be constrained by measuring the metallicity of the same cluster(s) with both instruments, at radii where scattered light is not a concern for \Suzaku{}. The best test case for such a direct comparison is the Coma cluster. Uniquely, Coma is a nearby, bright, non-centrally-peaked cluster that has been imaged out to large radii along several arms by \Suzaku{} \citep{Simionescu1302.4140} and has a very deep ($\sim1$\,Ms) \Chandra{} observation targeting radii close to $0.5\,r_{500}$.

Our analysis of this Coma data is detailed in Appendix~\ref{sec:coma}. In summary, the metallicity measured from \Chandra{} is lower than that measured by \Suzaku{} at the same radii in this cluster by $\sim40$ per cent, similar to the offset implied by the sample average metallicities in the 0.5--1.0\,$r_{500}$ aperture. This analysis motivates the prior $\ln(Z^\mathrm{Suz}/Z^\mathrm{Cha}) = 0.41\pm0.14$ marginalized over in our joint \Chandra{}+\Suzaku{} analysis in Section~\ref{sec:outskirts}. Although not definitive, the data are consistent with the possibility that charge transfer inefficiency (CTI) due to damage to the ACIS detectors aboard \Chandra{} reduces the flux observed in the Fe emission line, biasing the metallicity measurements low.

\section{Discussion} \label{sec:discussion}

\subsection{Comparison with Previous Work}

Since metallicity measurements for high-redshift clusters first became available, a number of authors have investigated the evolution of metallicity in the ICM. Table~\ref{tab:literature} lists results from the past $\sim10$ years where a power-law in $1+z$ was fitted to the data. There is a wide variation in the precise aperture adopted, the distribution of cluster redshifts, the analysis methods and the X-ray observatories used among these works, and correspondingly a range of results, with some indicating strong evolution in metallicity and some being consistent with a constant. Additional contributions beyond those listed in the table have been made by, e.g., \citet[][qualitatively consistent with a constant]{Leccardi0806.1445}, \citet[][qualitatively strong evolution]{Maughan0703156}, and \citet[][consistent with no evolution internally, but requiring strong evolution when combined with \citealt{Maughan0703156}]{Anderson0904.1007}. The most recent works are by \citet[][XMM data]{Ettori1504.02107}, who find evidence for evolution in the 0.0--0.15\,$r_{500}$ (0.15--0.4\,$r_{500}$) apertures at $\sim7\sigma$ ($\sim2\sigma$) significance, and consistency with a constant at radii $>0.4\,r_{500}$; and \citet[][\Chandra{} data]{McDonald1603.03035}, who find consistency with a constant value at $\sim1.5\sigma$ significance for a full 0.0--1.0\,$r_{500}$ aperture. 

\begin{table*}
  \begin{center}
    \caption{
      Summary of metallicity evolution constraints from the literature, where a power-law in $1+z$ is used to parametrized the evolution, for comparison to our results in Table~\ref{tab:fit}. [1] Reference; [2] redshift range of the cluster sample employed; [3] radial range used to fit the metallicity evolution in units of $r_{500}$; [4] reported constraint on the power-law index of metallicity as a function of $1+z$.
    }
    \label{tab:literature}
    \vspace{-1ex}
    \begin{tabular}{lccc}
      Paper & Redshifts & Aperture & $\beta_{1+z}$ \\
      \hline
      \citet{Balestra0609664} & 0.3--1.3 &  0.15--0.30 & $-1.25\pm0.15$ \\
      \citet{Baldi1111.4337} & 0.4--1.4 & 0.00--0.15 & \phmin$0.19^{+0.72}_{-0.68}$ \\
      \citet{Baldi1111.4337} & 0.4--1.4 & 0.15--0.40 & $-0.44^{+0.78}_{-0.72}$ \\
      \citet{Ettori1504.02107} & 0.09--1.4 & 0.00--0.15 & $-1.60\pm0.22$ \\
      \citet{Ettori1504.02107} & 0.09--1.4 & 0.15--0.40 & $-0.70\pm0.32$ \\
      \citet{Ettori1504.02107} & 0.09--1.4 & $>$0.4 & $-0.26\pm0.61$ \\
      \citet{McDonald1603.03035} & 0.25--1.5 & 0.0--1.0 & $-0.41\pm0.25$ \\
    \end{tabular}
  \end{center}
\end{table*}

\citet{Ettori1504.02107} find significant evolution in cluster centers ($r<0.15\,r_{500}$), driven by cool core clusters. When considering only non-cool core clusters, they find no evidence for evolution. Both of these observations are ostensibly at odds with our results; to compare them in greater depth, we would need to investigate in detail the correspondence between our H, M and L subsamples and the cool-core/non-cool-core division employed by \citet{Ettori1504.02107}. At intermediate radii, our 0.1--0.5\,$r_{500}$ evolution constraint is in good agreement with the 0.15--0.4\,$r_{500}$ result from \citet[][$\beta_{1+z}=-0.71\pm0.15$ vs.\ $-0.70\pm0.32$]{Ettori1504.02107}; in the outermost annulus considered, both works are consistent with a constant metallicity.

\citet{McDonald1603.03035} used \Chandra{} data for SPT selected clusters, most of which are also in our data set; indeed, nearly all of our clusters at $z>0.6$ are from SPT. Apart from the constraint listed in Table~\ref{tab:literature}, \citet{McDonald1603.03035} present their results on evolution (in 0.0--0.15 and 0.15--1\,$r_{500}$ apertures) in terms of a linear-in-redshift model which is not directly comparable to our fits. Their constraints on this linear model are compatible with a constant metallicity in both apertures. Given that measurements in the 0.15--1\,$r_{500}$ aperture should be most directly comparable to our 0.1--0.5\,$r_{500}$ results (due to the strong weighting of surface brightness towards smaller radii), the \citet{McDonald1603.03035} constraint at these radii is in conflict with our measurement of non-zero evolution in this aperture. However, as noted in Section~\ref{sec:intermediate}, our measurement of evolution is driven by the data at redshifts $z<0.4$, which are not well represented in an SPT-only data set (Figure~\ref{fig:sample}). A better comparison is with our data set at $z>0.4$ (or, almost equivalently, with only the SPT clusters in our data set), for which we find no evidence for evolution ($\beta_{1+z}=-0.23\pm0.56$; see Figure~\ref{fig:metal5}).

\subsection{The History of ICM Enrichment}

Measurements of metallicity in the Perseus and Virgo clusters revealed a uniform concentration of Fe outside of their cores and extending to their virial radii \citep{Werner1310.7948, Simionescu1704.01236}. In Virgo, the Si/Fe, S/Fe and Mg/Fe abundance ratios are also uniform at these radii \citep{Simionescu1506.06164}. Since enrichment of the ICM at late times by ram-pressure stripping or galactic outflows generically predicts a non-uniform metal distribution over this radial range (e.g.\ \citealt{Gunn1972ApJ...176....1G, Domainko0507605, Matsushita1301.0655}), and since large-scale mixing is prohibited by the steep entropy gradients in clusters, these observations have been taken as evidence that most of the metals in the ICM are produced prior to cluster formation. In this scenario, metals would need to be expelled from galaxies and mixed throughout the intergalactic medium in the proto-cluster environment; once accreted onto a cluster and virialized, significant further enrichment of the ICM can only occur in regions where the stellar density is relatively high (cluster cores). This early enrichment could have been driven by galactic winds \citep{De-Young1978ApJ...223...47D} powered by AGN and supernovae \citep{Madau9607172, Brandt0501058}. This picture is supported by the recent simulations of \citet{Fabjan0909.0664} and \citet{Biffi1701.08164}, who find that AGN feedback at $z\gtsim2$, when cluster potentials are still relatively shallow, is particularly effective at mixing metals throughout the ICM, producing similar radial trends to those observed, and minimal evolution.

The lack of metallicity evolution at radii $\gtsim0.5\,r_{500}$ measured by \citet{Ettori1504.02107} and in this work is consistent with early enrichment, as is the constant metallicity we find at somewhat smaller radii (0.1--0.5\,$r_{500}$) for redshifts $z>0.4$ (see also \citealt{McDonald1603.03035}). The late-time increase in metallicity that we measure in this intermediate radial range can be plausibly explained by small-scale mixing of the gas from cluster cores, where stellar evolution continues to enrich the ICM locally and where AGN feedback and cluster mergers provide a mechanism for small-scale mixing. While these observations produce a consistent picture of the history of enrichment, the precision of our metallicity constraints at high redshifts, especially at large radii, is extremely limited compared to what has been achieved in the nearby Universe with \Suzaku{}. A truly definitive test of early enrichment will require much more precise measurements of metallicity in the diffuse ICM, ideally at the highest redshifts possible. In the near term, one could envision doing this with deep XMM observations of high-$z$ clusters where the point source population has been constrained by \Chandra{}. In the future, observatories such as {\it Athena} or {\it Lynx}, with large collecting areas and high energy resolution, will revolutionize studies of the enrichment of the ICM at high redshifts.

\section{Conclusion} \label{sec:conclusion}

We have presented \Chandra{} measurements of the metallicity of the ICM in three radial ranges, 0.0--0.1, 0.1--0.5 and 0.5--1.0\,$r_{500}$, from a sample of 245 massive galaxy clusters. In the outermost radial range, we supplement this data set with recent, very precise measurements of nearby clusters from \Suzaku{} data. We fit these data with a model that includes power-law dependences on redshift and temperature, and distinguish between three classes of clusters (H, M and L) corresponding to different ranges of surface brightness peakiness. Our results can be summarized as follows.
\begin{itemize}
  \item In cluster centers (0.0--0.1\,$r_{500}$), metallicity clearly correlates with temperature and peakiness, in the sense of cooler and peakier clusters being more enriched. For the peakiest clusters (subsample H), we find a large intrinsic scatter and no overall trend with redshift, indicating that the lowest-entropy gas was enriched at early times, with variations in astrophysical processes such as feedback potentially driving the scatter. For the least-peaky (L) clusters, the scatter is small, and there is a trend of increasing metallicity with time. Medium-peakiness (M) clusters form an intermediate case.
  \item At intermediate radii (0.1--0.5\,$r_{500}$), we see evidence for evolution in the ICM metallicity, across morphological groups. However, this evolution appears to be limited to low redshifts, $z\ltsim0.4$. A plausible explanation is that, given enough time, this region can become contaminated by metals produced in cluster centers, aided by small-scale mixing due to gas sloshing or AGN activity.
  \item At larger radii (0.5--1.0\,$r_{500}$), the \Chandra{} data are consistent with a constant metallicity value for all clusters ($\beta_{1+z}=-0.30\pm0.91$). We combine these data with \Suzaku{} measurements of nearby clusters, accounting and fitting for a systematic offset between metallicities measured with the two telescopes. We find an average metallicity of $0.234\pm0.016$ relative to the solar abundances of \citet[][or $0.342\pm0.025$, correcting to the \Suzaku{} reference]{Asplund0909.0948}, and an intrinsic scatter consistent with zero ($<7$ per cent at 68.3 per cent confidence). Our constraint on the power-law dependence of metallicity on $1+z$ from these data, $\beta_{1+z}=0.35^{+1.18}_{-0.36}$, is consistent with zero and disfavors an increasing metallicity (ongoing enrichment) at these radii.
  \item Our results regarding the lack of evolution in metallicity at large radii are consistent with recent work by \citet{Ettori1504.02107} and \citet{McDonald1603.03035}. They are also consistent with the implication of the uniform metallicity in cluster outskirts measured in Perseus and Virgo \citep{Werner1310.7948, Simionescu1506.06164, Simionescu1704.01236}, namely that the ICM in these regions must have been enriched to its current level of $\sim0.3$ solar prior to cluster formation. Thereafter, the increases in metallicity that we find at smaller radii can be ascribed to stellar evolution products and the locally high stellar density in cluster centers. While this scenario is consistent with the current data, a definitive test of early enrichment would be best achieved through precise, spatially resolved measurements of metallicity in high-redshift clusters -- ideally at the highest redshifts where such observations are feasible.
\end{itemize}

\section*{Acknowledgements}
We are grateful to the anonymous referee for a number of useful comments and suggestions. We acknowledge support from the U.S. Department of Energy under contract number DE-AC02-76SF00515; from the National Aeronautics and Space Administration (NASA) under Grant No.\ NNX15AE12G, issued through the ROSES 2014 Astrophysics Data Analysis Program; and from NASA through Chandra Award Number GO6-17123X, issued by the Chandra X-ray Observatory Center, which is operated by the Smithsonian Astrophysical Observatory for and on behalf of NASA under contract NAS8-03060.

\def \aap {A\&A}
\def \aapr {A\&AR}
\def \apj {ApJ}
\def \apjl {ApJ}
\def \apjs {ApJS}
\def \araa {ARA\&A}
\def \asl {Adv.\ Sci.\ Lett.}
\def \mnras {MNRAS}
\def \nat {Nat}

\appendix

\section{Fitting Simulated Spectra} \label{sec:sims}

Here we show the results of testing our spectral fitting approach on simulated spectra. In particular, we contrast the use of the modified $C$ statistic with minimally binned spectra (grouped to $\geq1$ count per channel) with the practice of binning spectra more heavily ($\geq20$ counts per channel, to be concrete) and using the $\chi^2$ fit statistic (with ``standard'' weighting in {\sc xspec}, i.e. with the variance set equal to the number of observed counts in a bin). \citet{Leccardi0705.4199} and \citet{Humphrey0811.2796} present more detailed discussions of the relative performance of the two statistics; however, neither specifically addresses the case of \Chandra{} data and the associated practice of using blank-sky spectra to empirically model the background. In this appendix, we therefore consider that scenario for a cluster whose redshift and temperature are typical of our sample.

The starting point for our simulations is one of the intermediate-radius spectra for MACS\,J1115.8+0129. The redshift and absorbing column density, $z=0.355$ and $\NH=4.34\E{20}\cm^{-2}$, are kept fixed in our analysis.\footnote{Our results would certainly differ substantially if the cluster redshift needed to be simultaneously constrained from the X-ray data.} We adjust the best-fitting {\sc phabs$\ast$apec} model for this spectrum slightly, in the interest of working with round numbers, so that it has a temperature of 8\,keV, a metallicity of 0.3 solar, and a normalization that corresponds to 3000 net counts on average for an exposure time of 35\,ks (approximately the clean exposure of the original spectrum). Each realization consists of a Poisson draw from this model, along with the background, that is then grouped appropriately for the fit statistic being tested.

We analyzed $10^4$ Poisson realizations of the science and blank-sky spectra, for science exposures ranging from 4--70\,ks (nominally $\sim350$--7000 net counts). The simulated blank sky exposures were kept fixed at the actual exposure time of 400\,ks. For each realization, we found the model parameters that minimize the fit statistic, $S$ (where $S$ is either $C$ or $\chi^2$), and determined 1-dimensional 68.3 per cent confidence intervals on both temperature and metallicity based on the range of parameter values satisfying $\Delta S = S-S_\mathrm{min}\leq1$.

\begin{figure*}
  \centering
  \includegraphics[scale=\figscale]{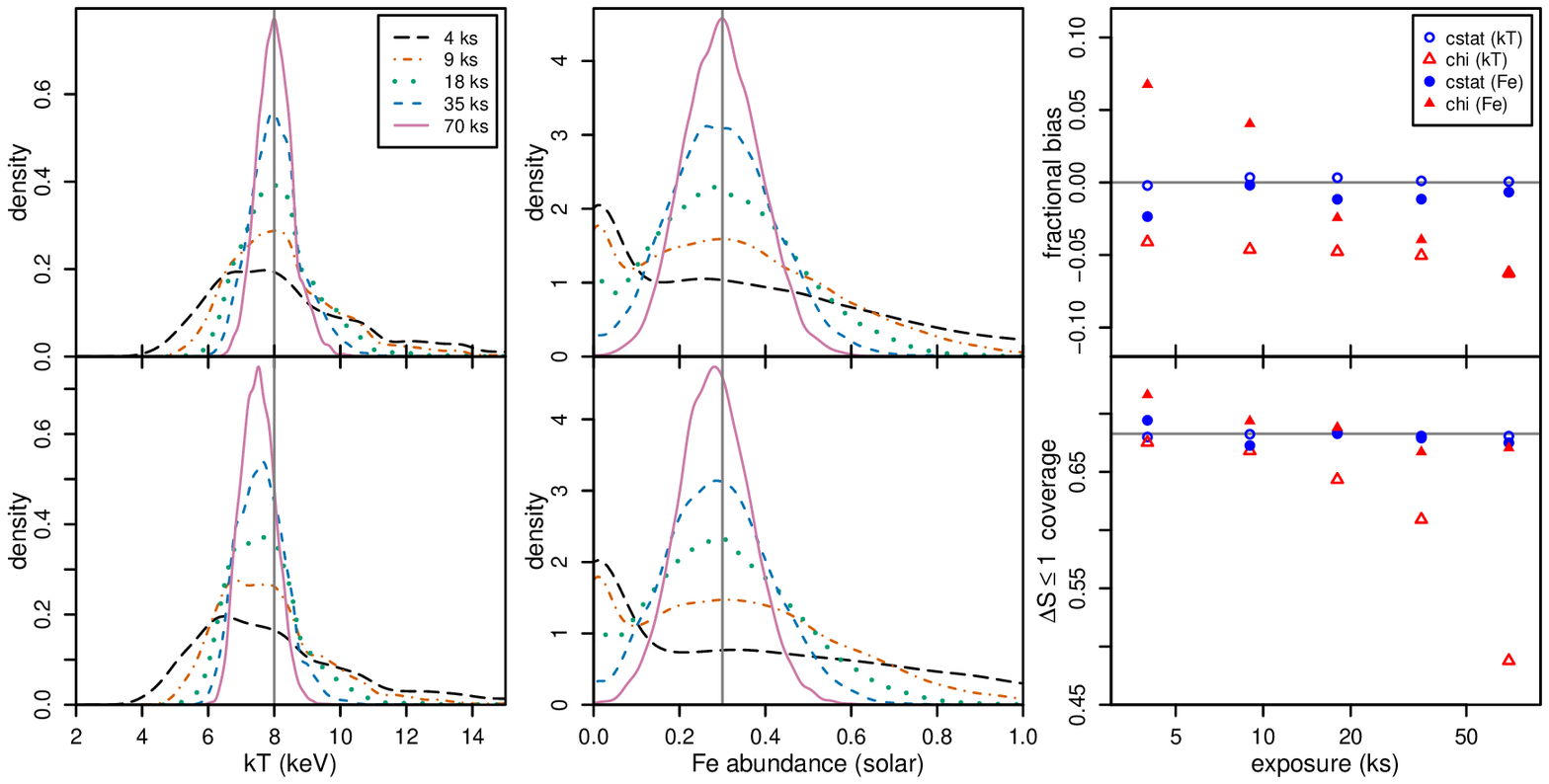}
  \caption{
    Left and center columns: smoothed histograms of the best-fitting temperature and metallicity values from many Poisson realizations of a model spectrum, fit using either the $C$-statistic (top row) or $\chi^2$ (bottom row), for various exposure times. The longest exposure of 70\,ks corresponds to 6000 source counts, on average.
    Right column: the fractional bias of the median of each histogram (top), and the corresponding frequentist coverages of the $1\sigma$ confidence intervals, defined by $\Delta C$ or $\Delta \chi^2\leq1$, as a function of exposure time.
  }
  \label{fig:sims}
\end{figure*}

Figure~\ref{fig:sims} summarizes the results of these simulations. The left and center columns show smoothed histograms of the best fitting values for various science exposure times, for temperature and metallicity (columns), and for each fit statistic ($C$ in the first row and $\chi^2$ in the second). The simulation input values are shown with vertical lines. The top-right panel shows the fractional bias of the median of each of those histograms as a function of simulated exposure time. The bottom-right panel shows the frequentist coverage of the $\Delta S\leq1$ confidence interval for each method, i.e.\ the fraction of simulations for a given method and exposure where the $\Delta S\leq1$ interval contains the true value, which should equal $\approx0.683$.\footnote{We have separately verified that the histograms in the figure agree well with the average Bayesian posterior distribution across realizations, as they should,  and similarly that our conclusions about confidence interval coverage obtained from $\Delta S$ agree with those derived from Bayesian posteriors. Note that the same priors, most significantly the requirement of non-negative metallicity, apply in both cases.}

We find that biases are minimal in the case of the $C$-statistic analysis, and that the coverage of the $1\sigma$ confidence intervals, constructed by requiring $\Delta C\leq1$, is accurate. In agreement with the authors cited above, we find that the same cannot be said of the $\chi^2$ analysis, even when the spectra are heavily binned and contain many thousand net counts. This is particularly the case for the temperature, for which the $\chi^2$ best fit is biased at the $\sim5$ per cent level for all the exposure times considered here, and for which the $\Delta \chi^2\leq1$ confidence intervals have significantly lower coverage than they should at the longest exposures.

Independent of the fit statistic, we also see the ``piling-up'' of probability at very low metallicities for low signal-to-noise spectra, as noted by \citet{Leccardi0705.4199}, resulting from the requirement that metallicity be non-negative. Our method for fitting Equation~\ref{eq:Zmodel}, which uses the full shape of the Bayesian posterior from each cluster, in principle accounts for both this effect and the generally asymmetric shape of many of the curves in Figure~\ref{fig:sims} (so long as the shape of the posterior is known sufficiently accurately). Methods which assume Gaussian or log-normal uncertainties, or indeed any fixed distribution \emph{centered about the best fit}, will clearly perform badly in this situation.

\section{Comparing Chandra and Suzaku Measurements of the Coma Cluster} \label{sec:coma}

In this appendix, we compare \Suzaku{} metallicity measurements of the Coma Cluster with those from a deep (963\,ks clean exposure), offset \Chandra{} observation (proposal ID 17800479). The data were reduced and analyzed in the same way as the rest of the \Chandra{} data used in this work.

\begin{figure*}
  \centering
  \includegraphics[scale=0.3]{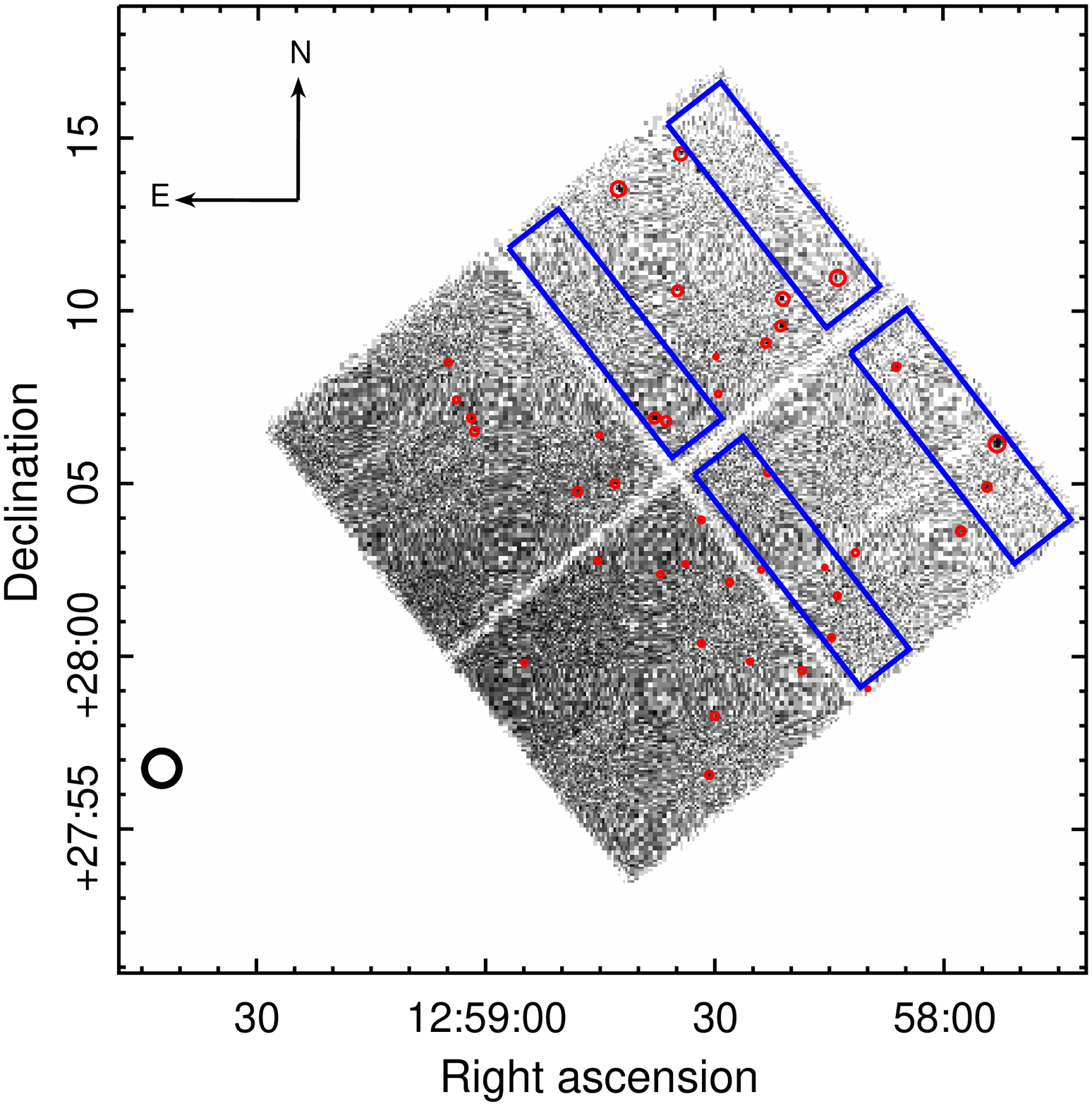}
  \hspace{1cm}
  \includegraphics[scale=\figscale]{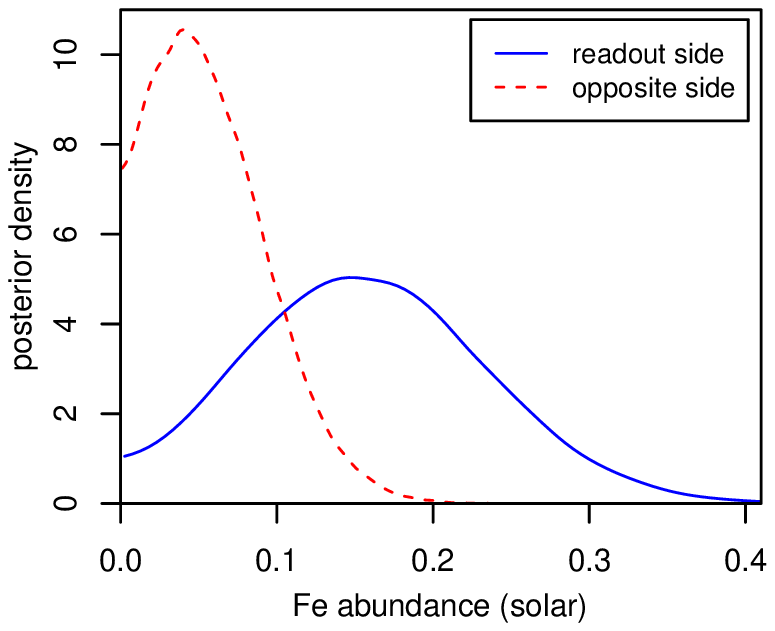}
  \caption{
    Left: representative \Chandra{} event map from the deep Coma data set (OBSID 18271; 53\,ks clean exposure). The $1'$ diameter black circle marks the cluster center adopted by \citet{Simionescu1302.4140}, while red circles indicate masked point sources. The blue rectangles overlaying the CCDs at larger cluster radii show regions adjacent to (upper right) and opposite (lower left) the readout strip. Within these CCDs, charge is transfered from the lower-left edge (closer to the cluster center) towards the upper-right edge during readout.
    Right: constraints on the metallicity of Coma from a subset of the \Chandra{} data. Blue solid and red dashed curves, respectively, correspond to readout-side and opposite-side regions of the CCDs farthest from the cluster center, as outlined in the left panel. Only observations with roll angles close to that shown in the left panel are used in these fits.
  }
  \label{fig:coma}
\end{figure*}

The field targeted by this observation lies west of the center of Coma, with the four ACIS-I CCDs typically covering cluster-centric radii of $\sim10'$--$25'$. This range spans the transition from a decreasing metallicity profile to an approximately flat one, as measured by \citet{Simionescu1302.4140} with \Suzaku{}. To conservatively exclude the decreasing part of the profile, we initially consider only the data at $r>13'$. Given the possibility that CTI may influence the metallicity measurements, we additionally restrict the analysis to CCDs which lie entirely at radii $>12'$, i.e.\  we do not use spectra extracted over partial CCDs. We note that the majority of the data (21/26 OBSIDs, or 77 per cent of the total clean exposure) was taken at roll angles similar to that shown in the left panel of Figure~\ref{fig:coma}, in which readout is approximately along the radial direction.

We fit the usual {\sc phabs$\ast$apec} model to the data identified above, linking the temperature and metallicity across all of the spectra, but allowing a different normalization for each. This results in a temperature of $8.6\pm0.07$\,keV, in good agreement with the \Suzaku{} measurements, and a metallicity of $0.189\pm0.024\Zsun$. Compared with the value of $0.283\pm0.018\Zsun$ measured from \Suzaku{} data at the same radii, this implies an offset of $\ln(Z^\mathrm{Suz}/Z^\mathrm{Cha}) = 0.41\pm0.14$, which we adopt as a prior in Section~\ref{sec:outskirts}.

If CTI is responsible for the reduced metallicity measured by \Chandra{}, we should see a dependence of the measurements on detector position: the metallicity measured from a region adjacent to the readout strip should be higher than that measured from the opposite side of the same CCD. To test this, we fit spectra extracted from the readout sides and the opposite sides of the two CCDs at larger cluster radii in the left panel of Figure~\ref{fig:coma}. Here we use only OBSIDs with roll angles similar to that shown (as noted above, this represents most of the data). At this roll angle, the readout strip for these CCDs lies at the far edge of ACIS-I from the cluster center. The signature of CTI, a higher metallicity measurement on the readout (large radius) side compared with the opposite (small radius) side, runs counter to our expectation that the true metallicity profile be flat (or, if anything, decreasing) at these radii. As shown in the right panel of Figure~\ref{fig:coma}, we do measure a higher metallicity on the readout side than on the opposite side ($0.15\pm0.07\Zsun$ compared with $0.04^{+0.04}_{-0.03}\Zsun$), although the uncertainties are relatively large. While we cannot necessarily conclude on this basis that CTI is responsible for the observed \Chandra-\Suzaku{} metallicity offset, the data are consistent with this hypothesis.

Since the effects of CTI have varied over the course of the \Chandra{} mission, and impact the front-illuminated (FI) and back-illuminated (BI) ACIS CCDs differently, metallicities measured from the archival data set could in principle also be used to test this hypothesis. Simple checks along these lines, splitting the spectra used in our archival analysis by detector (ACIS-S vs.\ ACIS-I) and by epoch of observation, were unable to detect trends within the significant statistical uncertainties. While it is beyond the scope of this work, it remains possible that a more intricate analysis, comparing spectra for the same regions of many multiply observed clusters while accounting for detector position and orientation, may shed more light on the issue. We note, however, that such a study would necessarily be very complex, given the complicated relationship that exists in practice between cluster mass and redshift, the date of observation (low-$z$, massive clusters preferentially earlier), the CCDs used (BI significantly more common early in the mission), and the placement of the cluster outskirts with respect to the active CCDs. In contrast, the Coma analysis above used exactly the same regions of the same CCDs, pointed at the same cluster gas with similar detector orientations, from observations spanning just over one year; in this case, even nearly 1\,Ms of exposure of one of the brightest clusters in the sky lacks the statistical power to clearly distinguish the detector regions studied.

% \bsp
\label{lastpage}

\begin{thebibliography}{}

\bibitem[\protect\citeauthoryear{{Allen}, {Evrard}, \& {Mantz}}{{Allen}
  et~al.}{2011}]{Allen1103.4829}
{Allen} S.~W., {Evrard} A.~E.,  {Mantz} A.~B., 2011,
  \href{http://adsabs.harvard.edu/abs/2011ARA%26A..49..409A}{\textcolor{blue}{\araa,
  49, 409}}

\bibitem[\protect\citeauthoryear{{Allen} \& {Fabian}}{{Allen} \&
  {Fabian}}{1998}]{Allen9802219}
{Allen} S.~W.,  {Fabian} A.~C., 1998,
  \href{http://adsabs.harvard.edu/abs/1998MNRAS.297L..63A}{\textcolor{blue}{\mnras,
  297, L63}}

\bibitem[\protect\citeauthoryear{{Allen} et~al.}{{Allen}
  et~al.}{2008}]{Allen0706.0033}
{Allen} S.~W., {Rapetti} D.~A., {Schmidt} R.~W., {Ebeling} H., {Morris} R.~G.,
  {Fabian} A.~C., 2008,
  \href{http://adsabs.harvard.edu/abs/2008MNRAS.383..879A}{\textcolor{blue}{\mnras,
  383, 879}}

\bibitem[\protect\citeauthoryear{{Anderson} et~al.}{{Anderson}
  et~al.}{2009}]{Anderson0904.1007}
{Anderson} M.~E., {Bregman} J.~N., {Butler} S.~C.,  {Mullis} C.~R., 2009,
  \href{http://adsabs.harvard.edu/abs/2009ApJ...698..317A}{\textcolor{blue}{\apj,
  698, 317}}

\bibitem[\protect\citeauthoryear{{Asplund} et~al.}{{Asplund}
  et~al.}{2009}]{Asplund0909.0948}
{Asplund} M., {Grevesse} N., {Sauval} A.~J.,  {Scott} P., 2009,
  \href{http://adsabs.harvard.edu/abs/2009ARA%26A..47..481A}{\textcolor{blue}{\araa,
  47, 481}}

\bibitem[\protect\citeauthoryear{{Baldi} et~al.}{{Baldi}
  et~al.}{2012}]{Baldi1111.4337}
{Baldi} A., {Ettori} S., {Molendi} S., {Balestra} I., {Gastaldello} F.,
  {Tozzi} P., 2012,
  \href{http://adsabs.harvard.edu/abs/2012A%26A...537A.142B}{\textcolor{blue}{\aap,
  537, A142}}

\bibitem[\protect\citeauthoryear{{Balestra} et~al.}{{Balestra}
  et~al.}{2007}]{Balestra0609664}
{Balestra} I., {Tozzi} P., {Ettori} S., {Rosati} P., {Borgani} S., {Mainieri}
  V., {Norman} C.,  {Viola} M., 2007,
  \href{http://adsabs.harvard.edu/abs/2007A%26A...462..429B}{\textcolor{blue}{\aap,
  462, 429}}

\bibitem[\protect\citeauthoryear{{Balucinska-Church} \&
  {McCammon}}{{Balucinska-Church} \&
  {McCammon}}{1992}]{Balucinska1992ApJ...400..699B}
{Balucinska-Church} M.,  {McCammon} D., 1992,
  \href{http://adsabs.harvard.edu/abs/1992ApJ...400..699B}{\textcolor{blue}{\apj,
  400, 699}}

\bibitem[\protect\citeauthoryear{{Biffi} et~al.}{{Biffi}
  et~al.}{2017}]{Biffi1701.08164}
{Biffi} V. et~al., 2017,
  \href{http://adsabs.harvard.edu/abs/2017MNRAS.468..531B}{\textcolor{blue}{\mnras,
  468, 531}}

\bibitem[\protect\citeauthoryear{{Bleem} et~al.}{{Bleem}
  et~al.}{2015}]{Bleem1409.0850}
{Bleem} L.~E. et~al., 2015,
  \href{http://adsabs.harvard.edu/abs/2015ApJS..216...27B}{\textcolor{blue}{\apjs,
  216, 27}}

\bibitem[\protect\citeauthoryear{{B{\"o}hringer} et~al.}{{B{\"o}hringer}
  et~al.}{2004}]{Bohringer0405546}
{B{\"o}hringer} H. et~al., 2004,
  \href{http://adsabs.harvard.edu/cgi-bin/nph-bib_query?bibcode=2004A%26A...425..367B&db_key=AST}{\textcolor{blue}{\aap,
  425, 367}}

\bibitem[\protect\citeauthoryear{{B{\"o}hringer} \& {Werner}}{{B{\"o}hringer}
  \& {Werner}}{2010}]{Boehringer2010A26ARv..18..127B}
{B{\"o}hringer} H.,  {Werner} N., 2010,
  \href{http://adsabs.harvard.edu/abs/2010A%26ARv..18..127B}{\textcolor{blue}{\aapr,
  18, 127}}

\bibitem[\protect\citeauthoryear{{Borgani} \& {Kravtsov}}{{Borgani} \&
  {Kravtsov}}{2011}]{Borgani0906.4370}
{Borgani} S.,  {Kravtsov} A., 2011,
  \href{http://adsabs.harvard.edu/abs/2009arXiv0906.4370B}{\textcolor{blue}{\asl,
  4, 204}}

\bibitem[\protect\citeauthoryear{{Brandt} \& {Hasinger}}{{Brandt} \&
  {Hasinger}}{2005}]{Brandt0501058}
{Brandt} W.~N.,  {Hasinger} G., 2005,
  \href{http://adsabs.harvard.edu/abs/2005ARA%26A..43..827B}{\textcolor{blue}{\araa,
  43, 827}}

\bibitem[\protect\citeauthoryear{{Cash}}{{Cash}}{1979}]{Cash1979ApJ...228..939}
{Cash} W., 1979,
  \href{http://adsabs.harvard.edu/abs/1979ApJ...228..939C}{\textcolor{blue}{\apj,
  228, 939}}

\bibitem[\protect\citeauthoryear{{De Grandi} et~al.}{{De Grandi}
  et~al.}{2004}]{De-Grandi0310828}
{De Grandi} S., {Ettori} S., {Longhetti} M.,  {Molendi} S., 2004,
  \href{http://adsabs.harvard.edu/abs/2004A%26A...419....7D}{\textcolor{blue}{\aap,
  419, 7}}

\bibitem[\protect\citeauthoryear{{De Young}}{{De
  Young}}{1978}]{De-Young1978ApJ...223...47D}
{De Young} D.~S., 1978,
  \href{http://adsabs.harvard.edu/abs/1978ApJ...223...47D}{\textcolor{blue}{\apj,
  223, 47}}

\bibitem[\protect\citeauthoryear{{Domainko} et~al.}{{Domainko}
  et~al.}{2006}]{Domainko0507605}
{Domainko} W. et~al., 2006,
  \href{http://adsabs.harvard.edu/abs/2006A%26A...452..795D}{\textcolor{blue}{\aap,
  452, 795}}

\bibitem[\protect\citeauthoryear{{Ebeling} et~al.}{{Ebeling}
  et~al.}{2007}]{Ebeling0703394}
{Ebeling} H., {Barrett} E., {Donovan} D., {Ma} C.-J., {Edge} A.~C.,  {van
  Speybroeck} L., 2007,
  \href{http://adsabs.harvard.edu/abs/2007ApJ...661L..33E}{\textcolor{blue}{\apjl,
  661, L33}}

\bibitem[\protect\citeauthoryear{{Ebeling} et~al.}{{Ebeling}
  et~al.}{1998}]{Ebeling1998MNRAS.301..881E}
{Ebeling} H., {Edge} A.~C., {Bohringer} H., {Allen} S.~W., {Crawford} C.~S.,
  {Fabian} A.~C., {Voges} W.,  {Huchra} J.~P., 1998,
  \href{http://adsabs.harvard.edu/cgi-bin/nph-bib_query?bibcode=1998MNRAS.301..881E&db_key=AST}{\textcolor{blue}{\mnras,
  301, 881}}

\bibitem[\protect\citeauthoryear{{Ebeling} et~al.}{{Ebeling}
  et~al.}{2010}]{Ebeling1004.4683}
{Ebeling} H., {Edge} A.~C., {Mantz} A., {Barrett} E., {Henry} J.~P., {Ma}
  C.~J.,  {van Speybroeck} L., 2010,
  \href{http://adsabs.harvard.edu/abs/2010MNRAS.407...83E}{\textcolor{blue}{MNRAS,
  407, 83}}

\bibitem[\protect\citeauthoryear{{Ettori} et~al.}{{Ettori}
  et~al.}{2015}]{Ettori1504.02107}
{Ettori} S., {Baldi} A., {Balestra} I., {Gastaldello} F., {Molendi} S.,
  {Tozzi} P., 2015,
  \href{http://adsabs.harvard.edu/abs/2015A%26A...578A..46E}{\textcolor{blue}{\aap,
  578, A46}}

\bibitem[\protect\citeauthoryear{{Fabjan} et~al.}{{Fabjan}
  et~al.}{2010}]{Fabjan0909.0664}
{Fabjan} D., {Borgani} S., {Tornatore} L., {Saro} A., {Murante} G.,  {Dolag}
  K., 2010,
  \href{http://adsabs.harvard.edu/abs/2010MNRAS.401.1670F}{\textcolor{blue}{\mnras,
  401, 1670}}

\bibitem[\protect\citeauthoryear{{Goodman} \& {Weare}}{{Goodman} \&
  {Weare}}{2010}]{Goodman10.2140-camcos.2010.5.65}
{Goodman} J.,  {Weare} J., 2010, {}{Communications in Applied Mathematics and
  Computational Science}, 5, 65

\bibitem[\protect\citeauthoryear{{Gunn} \& {Gott}}{{Gunn} \&
  {Gott}}{1972}]{Gunn1972ApJ...176....1G}
{Gunn} J.~E.,  {Gott} J.~R., III, 1972,
  \href{http://adsabs.harvard.edu/abs/1972ApJ...176....1G}{\textcolor{blue}{\apj,
  176, 1}}

\bibitem[\protect\citeauthoryear{{Humphrey}, {Liu}, \& {Buote}}{{Humphrey}
  et~al.}{2009}]{Humphrey0811.2796}
{Humphrey} P.~J., {Liu} W.,  {Buote} D.~A., 2009,
  \href{http://adsabs.harvard.edu/abs/2009ApJ...693..822H}{\textcolor{blue}{\apj,
  693, 822}}

\bibitem[\protect\citeauthoryear{{Kalberla} et~al.}{{Kalberla}
  et~al.}{2005}]{Kalberla0504140}
{Kalberla} P.~M.~W., {Burton} W.~B., {Hartmann} D., {Arnal} E.~M., {Bajaja} E.,
  {Morras} R.,  {P{\"o}ppel} W.~G.~L., 2005,
  \href{http://adsabs.harvard.edu/abs/2005A%26A...440..775K}{\textcolor{blue}{\aap,
  440, 775}}

\bibitem[\protect\citeauthoryear{{Leccardi} \& {Molendi}}{{Leccardi} \&
  {Molendi}}{2007}]{Leccardi0705.4199}
{Leccardi} A.,  {Molendi} S., 2007,
  \href{http://adsabs.harvard.edu/abs/2007A%26A...472...21L}{\textcolor{blue}{\aap,
  472, 21}}

\bibitem[\protect\citeauthoryear{{Leccardi} \& {Molendi}}{{Leccardi} \&
  {Molendi}}{2008}]{Leccardi0806.1445}
{Leccardi} A.,  {Molendi} S., 2008,
  \href{http://adsabs.harvard.edu/abs/2008A%26A...487..461L}{\textcolor{blue}{\aap,
  487, 461}}

\bibitem[\protect\citeauthoryear{{Madau} et~al.}{{Madau}
  et~al.}{1996}]{Madau9607172}
{Madau} P., {Ferguson} H.~C., {Dickinson} M.~E., {Giavalisco} M., {Steidel}
  C.~C.,  {Fruchter} A., 1996,
  \href{http://adsabs.harvard.edu/abs/1996MNRAS.283.1388M}{\textcolor{blue}{\mnras,
  283, 1388}}

\bibitem[\protect\citeauthoryear{{Mann} \& {Ebeling}}{{Mann} \&
  {Ebeling}}{2012}]{Mann1111.2396}
{Mann} A.~W.,  {Ebeling} H., 2012,
  \href{http://adsabs.harvard.edu/abs/2012MNRAS.420.2120M}{\textcolor{blue}{\mnras,
  420, 2120}}

\bibitem[\protect\citeauthoryear{{Mantz} et~al.}{{Mantz}
  et~al.}{2014}]{Mantz1402.6212}
{Mantz} A.~B., {Allen} S.~W., {Morris} R.~G., {Rapetti} D.~A., {Applegate}
  D.~E., {Kelly} P.~L., {von der Linden} A.,  {Schmidt} R.~W., 2014,
  \href{http://adsabs.harvard.edu/abs/2014MNRAS.440.2077M}{\textcolor{blue}{\mnras,
  440, 2077}}

\bibitem[\protect\citeauthoryear{{Mantz} et~al.}{{Mantz}
  et~al.}{2015a}]{Mantz1502.06020}
{Mantz} A.~B., {Allen} S.~W., {Morris} R.~G., {Schmidt} R.~W., {von der Linden}
  A.,  {Urban} O., 2015a,
  \href{http://adsabs.harvard.edu/abs/2015MNRAS.449..199M}{\textcolor{blue}{\mnras,
  449, 199}}

\bibitem[\protect\citeauthoryear{{Mantz} et~al.}{{Mantz}
  et~al.}{2016}]{Mantz1606.03407}
{Mantz} A.~B. et~al., 2016,
  \href{http://adsabs.harvard.edu/abs/2016MNRAS.463.3582M}{\textcolor{blue}{\mnras,
  463, 3582}}

\bibitem[\protect\citeauthoryear{{Mantz} et~al.}{{Mantz}
  et~al.}{2015b}]{Mantz1407.4516}
{Mantz} A.~B. et~al., 2015b,
  \href{http://adsabs.harvard.edu/abs/2015MNRAS.446.2205M}{\textcolor{blue}{\mnras,
  446, 2205}}

\bibitem[\protect\citeauthoryear{{Matsushita} et~al.}{{Matsushita}
  et~al.}{2013}]{Matsushita1301.0655}
{Matsushita} K., {Sakuma} E., {Sasaki} T., {Sato} K.,  {Simionescu} A., 2013,
  \href{http://adsabs.harvard.edu/abs/2013ApJ...764..147M}{\textcolor{blue}{\apj,
  764, 147}}

\bibitem[\protect\citeauthoryear{{Maughan} et~al.}{{Maughan}
  et~al.}{2008}]{Maughan0703156}
{Maughan} B.~J., {Jones} C., {Forman} W.,  {Van Speybroeck} L., 2008,
  \href{http://adsabs.harvard.edu/abs/2008ApJS..174..117M}{\textcolor{blue}{\apjs,
  174, 117}}

\bibitem[\protect\citeauthoryear{{McDonald} et~al.}{{McDonald}
  et~al.}{2016}]{McDonald1603.03035}
{McDonald} M. et~al., 2016,
  \href{http://adsabs.harvard.edu/abs/2016ApJ...826..124M}{\textcolor{blue}{\apj,
  826, 124}}

\bibitem[\protect\citeauthoryear{{Mushotzky} et~al.}{{Mushotzky}
  et~al.}{1978}]{Mushotzky1978ApJ...225...21M}
{Mushotzky} R.~F., {Serlemitsos} P.~J., {Boldt} E.~A., {Holt} S.~S.,  {Smith}
  B.~W., 1978,
  \href{http://adsabs.harvard.edu/abs/1978ApJ...225...21M}{\textcolor{blue}{\apj,
  225, 21}}

\bibitem[\protect\citeauthoryear{{Planck Collaboration}}{{Planck
  Collaboration}}{2016}]{Planck1502.01598}
{Planck Collaboration}, 2016,
  \href{http://adsabs.harvard.edu/abs/2016A%26A...594A..27P}{\textcolor{blue}{\aap,
  594, A27}}

\bibitem[\protect\citeauthoryear{{Santos} et~al.}{{Santos}
  et~al.}{2012}]{Santos1111.3642}
{Santos} J.~S., {Tozzi} P., {Rosati} P., {Nonino} M.,  {Giovannini} G., 2012,
  \href{http://adsabs.harvard.edu/abs/2012A%26A...539A.105S}{\textcolor{blue}{\aap,
  539, A105}}

\bibitem[\protect\citeauthoryear{{Simionescu} et~al.}{{Simionescu}
  et~al.}{2017}]{Simionescu1704.01236}
{Simionescu} A., {Werner} N., {Mantz} A., {Allen} S.~W.,  {Urban} O., 2017,
  \href{http://adsabs.harvard.edu/abs/2017MNRAS.469.1476S}{\textcolor{blue}{\mnras,
  469, 1476}}

\bibitem[\protect\citeauthoryear{{Simionescu} et~al.}{{Simionescu}
  et~al.}{2013}]{Simionescu1302.4140}
{Simionescu} A. et~al., 2013,
  \href{http://adsabs.harvard.edu/abs/2013ApJ...775....4S}{\textcolor{blue}{\apj,
  775, 4}}

\bibitem[\protect\citeauthoryear{{Simionescu} et~al.}{{Simionescu}
  et~al.}{2015}]{Simionescu1506.06164}
{Simionescu} A., {Werner} N., {Urban} O., {Allen} S.~W., {Ichinohe} Y.,
  {Zhuravleva} I., 2015,
  \href{http://adsabs.harvard.edu/abs/2015ApJ...811L..25S}{\textcolor{blue}{\apjl,
  811, L25}}

\bibitem[\protect\citeauthoryear{{Th{\"o}lken} et~al.}{{Th{\"o}lken}
  et~al.}{2016}]{Tholken1603.05255}
{Th{\"o}lken} S., {Lovisari} L., {Reiprich} T.~H.,  {Hasenbusch} J., 2016,
  \href{http://adsabs.harvard.edu/abs/2016A%26A...592A..37T}{\textcolor{blue}{\aap,
  592, A37}}

\bibitem[\protect\citeauthoryear{{Urban} et~al.}{{Urban}
  et~al.}{2017}]{Urban1706.01567}
{Urban} O., {Werner} N., {Allen} S.~W., {Simionescu} A.,  {Mantz} A., 2017,
  \href{http://adsabs.harvard.edu/abs/2017MNRAS.470.4583U}{\textcolor{blue}{\mnras,
  470, 4583}}

\bibitem[\protect\citeauthoryear{{Werner} et~al.}{{Werner}
  et~al.}{2013}]{Werner1310.7948}
{Werner} N., {Urban} O., {Simionescu} A.,  {Allen} S.~W., 2013,
  \href{http://adsabs.harvard.edu/abs/2013Natur.502..656W}{\textcolor{blue}{\nat,
  502, 656}}

\end{thebibliography}
\end{document}